\documentclass[%
preprint,
showpacs,
amsmath,amssymb,
aps,
prb,
]{revtex4-1}
\usepackage{graphicx}
\usepackage{dcolumn}
\usepackage{bm}
\usepackage{color}
\usepackage{verbatim}

\begin{document}

\preprint{APS/123-QED}

\title{Effects of dynamical paths on the energy gap and the corrections to free energy in path integrals of mean-field quantum spin systems}

\author{Yang Wei Koh}
\email{patrickkyw@gmail.com}
\affiliation{Bioinformatics Institute, 30 Biopolis Street, No. 07-01, Matrix, Singapore 138671}

\date{\today}

\begin{abstract}

In current studies of mean-field quantum spin systems, much attention is placed on the calculation of the ground-state energy and the excitation gap, especially the latter which plays an important role in quantum annealing. In pure systems, the finite gap can be obtained by various existing methods such as the Holstein-Primakoff transform, while the tunneling splitting at first-order phase transitions has also been studied in detail using instantons in many previous works. In disordered systems, however, it remains challenging to compute the gap of large-size systems with specific realization of disorder. Hitherto, only quantum Monte Carlo techniques are practical for such studies. Recently, Knysh [Nature Comm. \textbf{7}, 12370 (2016)] proposed a method where the exponentially large dimensionality of such systems is condensed onto a random potential of much lower dimension, enabling efficient study of such systems. Here we propose a slightly different approach, building upon the method of static approximation of the partition function widely used for analyzing mean-field models. Quantum effects giving rise to the excitation gap and non-extensive corrections to the free energy are accounted for by incorporating dynamical paths into the path integral. The time-dependence of the trace of the time-ordered exponential of the effective Hamiltonian is calculated by solving a differential equation perturbatively, yielding a finite-size series expansion of the path integral. Formulae for the first excited-state energy are proposed to aid in computing the gap. We illustrate our approach using the infinite-range ferromagnetic Ising model and the Hopfield model, both in the presence of a transverse field.

\end{abstract}
\maketitle


\section{Introduction}
\label{sec.I.intro}

The study of quantum spin systems is currently receiving much attention in various fields such as quantum annealing\cite{Liu15,Knysh16,Matsuura17}, quantum spin liquids\cite{Balents10,Savary17}, and machine learning\cite{Carrasquilla17}. The calculation of the partition function via the path integral is one of the major approaches for analyzing such systems. One early formulation is the spin coherent state path integral\cite{Klauder79,Kuratsuji80,Kuratsuji81} of which the semiclassical propagator incorporating the Solari-Kochetov phase has been derived by various authors\cite{Kochetov95,Stone00} and applied to isolated spins and homogeneous systems\cite{Garg03}. For many-body systems, the spin-spin interactions are usually decoupled by introducing auxiliary Hubbard-Stratonovich fields, reducing the calculation of the path integral to that of the trace of a time-ordered exponential of an effective single-body Hamiltonian\cite{Bray80,Thirumalai89,Nishimori96,Jorg10,Bapst12,Galitski11,Ringel13}. One development involves the treatment of the single-spin trace using Lie-algebraic methods. Galitski formulated the trace in terms of the time evolution of a density matrix on a Lie group and solved the operator equation of motion for a number of Lie algebras\cite{Galitski11}. Ringel and Gritsev pointed out that the product of time-ordered exponentials is equivalent to an ordinary product of exponentials where the time-orderedness has been `disentangled'\cite{Ringel13}. They showed that the disentangling conditions take the form of the Riccati equation, and applied their method to the one-dimensional Ising chain and an atom interacting with a photonic waveguide. On the numerical front, efficient quantum Monte Carlo algorithms have also been developed catering to both short-range lattice systems\cite{Sandvik10} as well as mean-field type models\cite{Young08,Young10}.

In a recent development, the effects of disorder in quantum spin models is studied by Knysh via the partition function of the two-pattern Gaussian Hopfield model\cite{Knysh16}. By transforming to the instantaneous adiabatic representation, the author recasts a many-body path integral into an equivalent single-body one where the effective Lagrangian describes an ordinary quantum mechanical particle with the familiar kinetic and potential energy terms. An interesting insight gained from this reformulation is that the complexities of the original disordered many-body problem are now compactly summarized in form of a random potential, and one can solve the one-particle quantum mechanical problem instead of computing the partition function. With this approach, the author is able to compute the energy spectrum describing the quantum annealing of large-size systems with specific realizations of disorder. 

In addition to the partition function, another important quantity in the study of quantum spin systems is the energy gap between the ground and first-excited states. In one-dimensional systems, the gap can usually be obtained using the Jordan-Wigner transform\cite{Lieb61,Okuyama15}. Garg et al. computed the instanton approximation of the spin coherent state path integral and obtained the tunneling splitting of the Lipkin-Meshkov-Glick model and the molecular magnetic Fe$_8$ in a transverse field\cite{Garg03}. For certain so-called integrable systems, the operator-based approach combining the Holstein-Primakoff transform and continuous unitary transformations has been used to obtain the $1/N$ expansion of the energy gap\cite{Dusuel04,Dusuel05}. In quantum annealing, much attention has been paid to the finite-size scaling of the gap at a phase transition point as the annealing success rate depends on the minimum gap along the annealing trajectory\cite{Suzuki07,Ohzeki11}. In a detailed study of the ferromagnetic $p$-spin model, J\"{o}rg et al. calculated both the finite and closing gaps numerically as well as analytically using various methods such as Rayleigh-Schr\"{o}dinger perturbation theory and instantons\cite{Jorg10}. A similar approach was recently used and extended to finite temperature in the study of quantum annealing correction by Matsuura et al.\cite{Matsuura17}. In disordered systems, the calculation of the gap is complicated by the presence of quenched random variables in the Hamiltonian. In exact numerical studies, the full Hamiltonian matrix is diagonalized and hence results are limited to small system sizes\cite{Hogg03,Jorg08,Takahashi10a,Takahashi10b}. The tunneling splitting in the quantum random energy model has also been obtained by first averaging over the disorder using the replica trick and then applying instanton calculus to the disorder-averaged static free energy\cite{Jorg08}. In actual applications of quantum annealing, however, the gap always depends on specific realizations of disorder. The gap of individual realizations has been calculated by Young et al. in the context of the quantum exact cover problem using quantum Monte Carlo simulations\cite{Young08,Young10}. The statistical properties of small energy gaps in the glassy phase of disordered systems and the implications for quantum annealing has also been examined recently by Knysh using the random potential formulation reviewed above\cite{Knysh16}.

In this work, we focus on a particular group of mean-field models where the inter-spin interactions can be expressed in the form $\left(\sum_i x_i\sigma_i\right)^l$ where $\sigma_i$ is the state of the $i$th spin, $l$ is a positive integer, and $x_i$ are parameters of the system. Although this consists of only a small subgroup of possible spin Hamiltonians, it still covers a reasonable range of ordered\cite{Jorg10,Bapst12,Matsuura17,Seki12,Seoane12} and disordered\cite{Knysh16,Nishimori96,Seki15} models, some of interest in studies of quantum annealing. We follow the popular approach of treating the partition function, first using the Suzuki-Trotter decomposition to perform a mapping onto a classical model with an additional `time' dimension, and then introducing auxiliary fields to decouple the inter-spin interactions. The path integral of the partition function $Z$ then takes a general form
\begin{equation} 
Z
=
\int \mathcal{D}m(t)
\exp
\left\{
-\beta N F[m(t)]
\right\},
\label{eq.sec_I.01}
\end{equation} 
where $\beta$ is the inverse temperature, $N$ is the total number of spins, and $F[m(t)]$ is a functional of the auxiliary field $m(t)$ introduced to decouple the interactions $\left(\sum_i x_i\sigma_i\right)^l$. The time-dependent $m(t)$ also serves as the path and is sometimes interpreted as an order parameter (e.g. magnetization). $\int \mathcal{D}m(t)$ denotes summing over all possible paths. $F[m(t)]$ contains $\mathcal{T}$, the trace  of the time-ordered exponential of the effective Hamiltonian $\mathcal{H}^{\mathrm{eff}}$ involving variables of a single spin. Denoting discretized time as $\kappa$, 
\begin{equation}
\mathcal{T}
=
\mathrm{Tr}
\left[
\prod_{\kappa}
\exp
\left[
\mathcal{H}^{\mathrm{eff}}
[m(\kappa)]
\cdot
\Delta\kappa
\right]
\right]
,
\label{eq.sec_I.05}
\end{equation}
where `Tr' denotes taking trace and $\Delta\kappa$ is an infinitesimal time interval. $\mathcal{H}^{\mathrm{eff}}$ depends on the value of the path $m(\kappa)$ at time $\kappa$. If $m(\kappa)$ varies with $\kappa$, then $\mathcal{H}^{\mathrm{eff}}$ at different times do not commute, making it difficult to evaluate the infinite product. Note that in the spin coherent state propagator\cite{Klauder79,Kuratsuji80,Kuratsuji81,Kochetov95,Stone00,Garg03}, one does not encounter $\mathcal{T}$ because auxiliary fields are not introduced to decouple the inter-spin interactions. Similarly, working in the instantaneous adiabatic representation, Knysh obtained in place of $\mathcal{T}$ a kinetic energy term in the effective Lagrangian\cite{Knysh16}. On the other hand, as mentioned above, $\mathcal{T}$ is also the central object in various Lie-algebraic methods\cite{Galitski11,Ringel13}. These works, however, have focused mainly on lattice systems where the dimension of the auxiliary fields is same as the total number of spins, and the path integration is different from the mean-field models we consider here. For Eq. (\ref{eq.sec_I.01}), when studying equilibrium thermodynamic properties, one usually makes the so-called static approximation that $m(t)$ is constant in time\cite{Jorg10,Bapst12,Matsuura17,Nishimori96,Seki15}. $\mathcal{T}$ and $F$ becomes time-independent, and Eq. (\ref{eq.sec_I.01}) is evaluated using the method of steepest descent.

\begin{figure}[h]
\begin{center}
\includegraphics[scale=0.9]{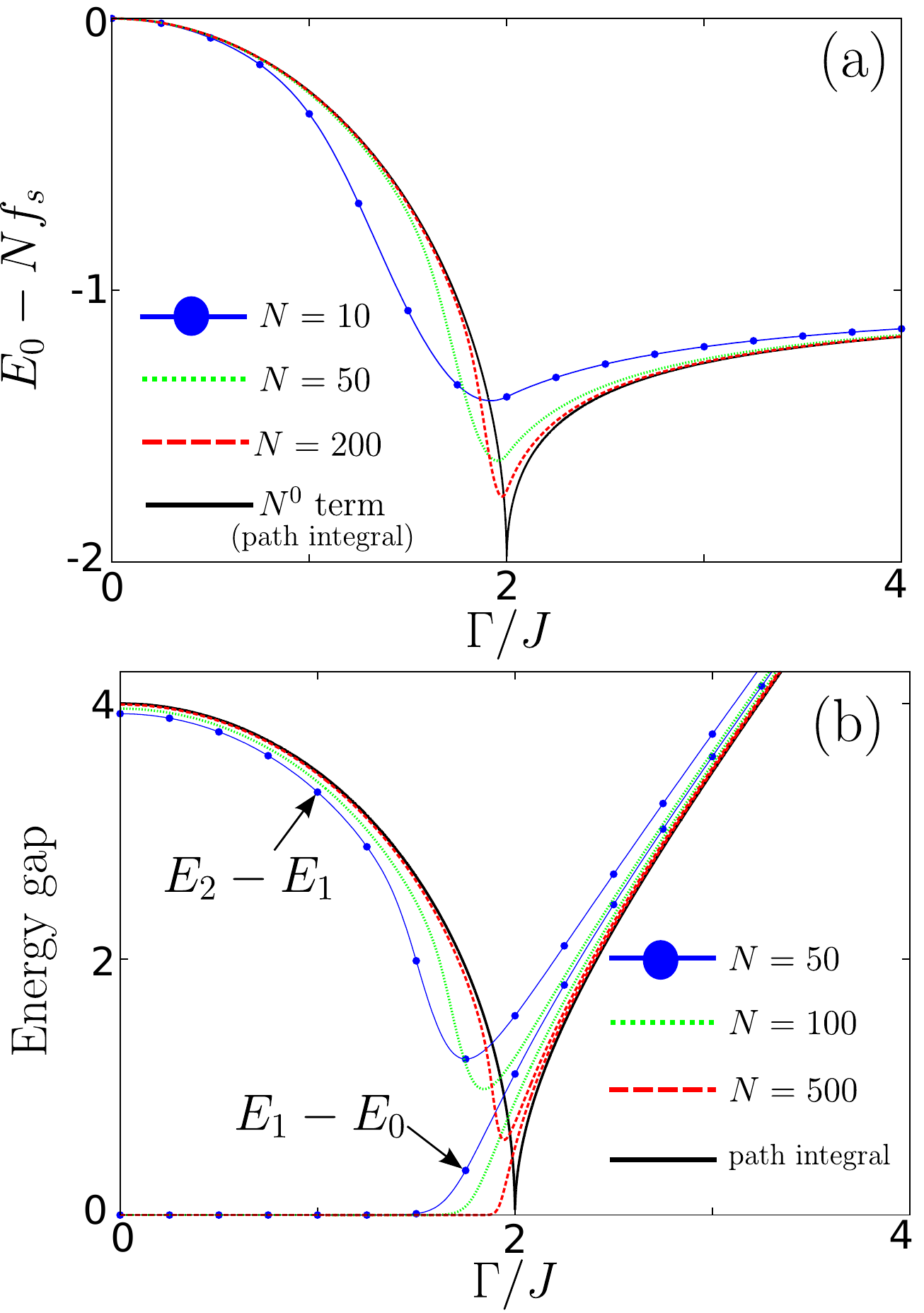}
\caption{Results of numerical diagonalization of the ferromagnetic model in the sector with total angular momentum $N/2$. (a) Difference between the ground-state energy $E_0$ and the free energy $Nf_s$ for various $N$. The $N^0$ term (black solid line) is the first correction to the free energy obtained by incorporating dynamical paths into the path integral [Eq. (\ref{eq.sec_III.04})]. (b) The energy gap for various $N$. For each $N$, the upper curve is $E_2-E_1$ while the lower one is $E_1-E_0$, as indicated for the case of $N=50$. In the ferromagnetic phase ($\Gamma<2J$), the ground-state is doubly degnerate and $E_2$ is the energy of the first-excited state. The curve labelled `path integral' (black solid line) is the gap obtained using our path integral formulation of the energy gap.}
\label{fig.fig01}
\end{center}
\end{figure}

In the static approximation, only the leading extensive (i.e. $\propto N$) contribution to the integral Eq. (\ref{eq.sec_I.01}) is retained while higher-order terms are neglected. To illustrate, consider the infinite-range ferromagnetic Ising model in a transverse field given by the Hamiltonian
\begin{equation}
H=-\frac{J}{N}\left(\sum_{i=1}^{N}\sigma_i^{z}\right)^2 -\Gamma \sum_{i=1}^{N}\sigma_i^x,
\label{eq.sec_I.03}
\end{equation}
where $\sigma_i^{\mu}$ $(\mu=x,y,z)$ is the $\mu$-direction Pauli matrix of the $i$th spin, and $J$ and $\Gamma$ are, respectively, the strengths of the ferromagnetic coupling and the transverse field. The ground-state energy $E_0$ is obtained from $Z$ via the relation,
\begin{equation}
E_0=\lim_{\beta\rightarrow \infty}-\frac{1}{\beta}\ln Z
.
\label{eq.sec_I.02}
\end{equation}
We computed the exact ground-state energy of Eq. (\ref{eq.sec_I.03}) by numerical diagonalization, and the free energy by substituting the static approximation of $Z$ into Eq. (\ref{eq.sec_I.02}). The difference between the two is shown in Fig. \ref{fig.fig01}(a) for several $N$. We see that the error is of order $N^0$ and quite significant. To go beyond the free energy, it is necessary to calculate higher-order terms in the $N^{-1}$ expansion of $E_0$. Improving upon the order $N^1$ accuracy of static approximation is also important when dealing with the energy gap. As the gap arises from the excitation of just a few spins, its magnitude is of order $N^0$ or even smaller, so static approximation is too coarse to evaluate even the leading order of the gap. To illustrate again, we computed the energy gap of Eq. (\ref{eq.sec_I.03}) numerically for various $N$ and the results are shown in Fig. \ref{fig.fig01}(b). It is seen from the scale of the vertical axis that the size of the gap is of the same order of magnitude as the error between $E_0$ and the free energy shown in Fig. \ref{fig.fig01}(a). Apart from resolving the gap, terms in the higher-order expansion of the path integral have other applications. For instance, the coefficient of the $N^{-1}$ term in the expansion of $E_0$ contains information about entanglement and is related to the scaling exponents of the correlation functions of finite-size systems\cite{Dusuel04,Dusuel05}. In disordered mean-field systems, the landscape of the effective potential energy is rugged on an energy scale with magnitude $N^0$, so one needs to go beyond the static approximation to be able to resolve small energy gaps within the spin glass phase\cite{Knysh16}.

In this paper, we consider the path integral of mean-field quantum spin systems where the functional $F$ in Eq. (\ref{eq.sec_I.01}) is formulated in terms of the single-spin trace $\mathcal{T}$ given by Eq. (\ref{eq.sec_I.05}). Our work has two objectives. The first is to improve upon the static approximation of $Z$ by incorporating dynamical fluctuations into the paths as 
\begin{equation}
m(t)=m_s + \lambda m_d(t),
\label{eq.sec_I.04}
\end{equation}
where the static part $m_s$ is weakly perturbed by the dynamical part $m_d(t)$ and $\lambda=1/\sqrt{N}$ is the strength of perturbation. One difficulty with incorporating Eq. (\ref{eq.sec_I.04}) is that substituting it into the trace $\mathcal{T}$, the expansion in powers of $\lambda$ is not straightforward. We propose mapping $\mathcal{T}$ onto a time-dependent ordinary differential equation governing the evolution of the state of a single spin. Consider
\begin{equation}
\frac{d|\psi(t)\rangle}{dt}
=
\mathcal{H}^{\mathrm{eff}}[m(t)]
|\psi(t)\rangle
,
\label{eq.sec_I.06}
\end{equation}
where the spin state $|\psi(t)\rangle$ is a two-component spinor at time $t$. The infinite product in Eq. (\ref{eq.sec_I.05}) is actually the fundamental matrix of the differential equation Eq. (\ref{eq.sec_I.06}). Hence, $\mathcal{T}$ can be calculated by solving Eq. (\ref{eq.sec_I.06}). Furthermore, with Eq. (\ref{eq.sec_I.04}), $\mathcal{H}^{\mathrm{eff}}$ becomes 
\begin{equation}
\mathcal{H}^{\mathrm{eff}}[m(t)]=\mathcal{H}_s[m_s] + \lambda \mathcal{H}_d[m_d(t)]
,
\label{eq.sec_I.07}
\end{equation}
where the static part $\mathcal{H}_s$ is perturbed by a small time-dependent term $\mathcal{H}_d$, so Eq. (\ref{eq.sec_I.06}) can be solved perturbatively. The solution enables us to expand $\mathcal{T}$ and $F[m(t)]$ in powers of $N^{-1}$, thereby evaluating the path integral beyond the extensive term. 

Our second contribution concerns the energy gap, specifically focusing on its formulation in terms of path integrals. The tunneling splitting between two degenerate energy minima\cite{Garg03} and at a first-order phase transition\cite{Matsuura17,Jorg10,Jorg08} has received much attention in the literature using the established instanton method. Recently, a more general method that can calculate both tunneling splitting as well as the finite gap in any phase was proposed where, instead of evaluating the path integral explicitly, one works directly with the random potential in the effective Lagrangian\cite{Knysh16}. Here, we focus on calculating the finite gap by directly evaluating the path integral. In Eq. (\ref{eq.sec_I.02}), $E_0$ appears on the left hand side of the relation. To obtain information about the energy gap, one must also calculate the first excited-state energy $E_1$. To this end, recall that
\begin{equation}
Z
=
\mathrm{Tr}
\left(
e^{-\beta H}
\right)
=
e^{-\beta E_0}
+
e^{-\beta E_1}
+
\cdots
,
\label{eq.sec_I.08}
\end{equation}
where we assume that the energy levels are non-degenerate. Consider another function $Z_Q$ defined as 
\begin{equation}
Z_Q
=
\mathrm{Tr}
\left(
Q e^{-\beta H}
\right)
=
\langle E_0|Q|E_0\rangle
e^{-\beta E_0}
+
\langle E_1|Q|E_1\rangle
e^{-\beta E_1}
+
\cdots
,
\label{eq.sec_I.09}
\end{equation}
where $Q$ is an appropriate operator to be defined shortly. Suppose a $Q$ can be found such that $\langle E_0|Q|E_0\rangle=1=-\langle E_1|Q|E_1\rangle$; subtracting Eq. (\ref{eq.sec_I.09}) from (\ref{eq.sec_I.08}), the leading term $e^{-\beta E_0}$ disappears and one gets
\begin{equation}
E_1=\lim_{\beta\rightarrow \infty}-\frac{1}{\beta}\ln \left(Z-Z_{Q}\right)
.
\label{eq.sec_I.10}
\end{equation}
From Eq. (\ref{eq.sec_I.10}), the gap is obtained from $E_1-E_0$. The right hand side of the relation is to be evaluated using the dynamical path integral approach described earlier. 

The above is a specific case of our proposed strategy for formulating the energy gap in a way amenable to path integral treatments. More generally, one considers
\begin{equation}
\mathrm{Tr}
\left[
f\left(Q,e^{-\beta H}\right)
\right]
=
f_0(Q)
e^{-\beta E_0}
+
f_1(Q)
e^{-\beta E_1}
+
\cdots
,
\label{eq.sec_I.11}
\end{equation}
where $Q$ and $f$ are operator and function that are to be chosen such that the coefficient $f_0(Q)$ vanishes while $f_1(Q)$ remains non-zero, making $e^{-\beta E_1}$ instead of $e^{-\beta E_0}$ the leading term. The choice of $Q$ and $f$ are system dependent, but one can make guesses based on the symmetry of the model. For concreteness, consider the choice of $Q$ in Eq. (\ref{eq.sec_I.09}) for the ferromagnetic model Eq. (\ref{eq.sec_I.03}). The model displays overall spin-flip symmetry, and we can let $Q$ be the parity operator
\begin{equation}
Q=\prod_{i=1}^{N}\sigma_i^x
.
\label{eq.sec_I.12}
\end{equation}
As $Q$ commutes with $H$, the two lowest energy levels are parity eigenstates: $Q|E_0\rangle=|E_0\rangle$ and $Q|E_1 \rangle=-|E_1 \rangle$. Eq. (\ref{eq.sec_I.10}) is therefore applicable to the model Eq. (\ref{eq.sec_I.03}). One limitation of Eq. (\ref{eq.sec_I.10}), however, is that the relation is applicable only in the paramagnetic phase of Eq. (\ref{eq.sec_I.03}). In Sec. \ref{sec.V.gap in both phases}, we propose a second relation which is applicable to the paramagnetic as well as the ferromagnetic phase. 

In the following we first illustrate the ideas presented above using the ferromagnetic model Eq. (\ref{eq.sec_I.03}). To demonstrate that our approach has broader applicability, we then apply parts of our methods to a disordered model, the Hopfield model in a transverse field\cite{Nishimori96}. This model has recently received attention within the static approximation in the context of quantum annealing\cite{Seki15,MatsuuraPRL17}. A related system, the Gaussian Hopfield model, was recently examined beyond the static approximation by Knysh using a slightly different formulation\cite{Knysh16}. Although our results here are also obtainable by that of Knysh, as reviewed above, we follow a different formulation of the path integral. In our analysis of the Hopfield model, the physical quantities that we investigated are also slightly different. Some interesting features of disordered systems elucidated by Knysh that we recover in our work will be discussed.

The rest of the paper is organized as follows. In Sec. \ref{sec.II.expansions of T and Z}, we derive the perturbative expansions of $\mathcal{T}$ and $Z$. Secs. \ref{sec.III.N0 and N-1 terms of E0} to \ref{sec.V.gap in both phases} concern the ferromagnetic model. In Sec. \ref{sec.III.N0 and N-1 terms of E0} we calculate the $N^0$ and $N^{-1}$ terms of $E_0$. In  Sec. \ref{sec.IV.ZQ} we calculate $Z_Q$ and use Eq. (\ref{eq.sec_I.10}) to obtain the energy gap in the paramagnetic phase. In Sec. \ref{sec.V.gap in both phases} we derive a trace formula for $E_0+E_1$ analogous to Eqs. (\ref{eq.sec_I.02}) and (\ref{eq.sec_I.10}) and use it to calculate the gap in both phases. Sec. \ref{sec.2nd_reply_PRB.Hopfield_Model} concerns the Hopfield model. After a presenting a careful treatment of disorder in the static approximation, the methods of Secs. \ref{sec.II.expansions of T and Z} to \ref{sec.IV.ZQ} are applied to calculate the ground-state energy and various energy gaps. Sec. \ref{sec.VI.conclusion} discusses and concludes the paper.

\section{Perturbative expansions of $\mathcal{T}$ and $Z$}
\label{sec.II.expansions of T and Z}

The path integral of $Z$ is obtained via the Suzuki-Trotter decomposition
\begin{eqnarray}
Z&=&\lim_{M\rightarrow \infty} Z_M \nonumber\\
 &=& \lim_{M\rightarrow \infty} \mathrm{Tr}
\left(
\left[
e^{\frac{\beta J}{M N } \left(\sum_i \sigma_i^z\right)^2}
e^{\frac{\beta \Gamma}{M}\sum_i \sigma_i^x}
\right]^{M}
\right).
\label{eq.sec_II.01}
\end{eqnarray}
Resolutions of identity in the $z$-basis are inserted between each pair of exponentials to evaluate $\left(\sum_i \sigma_i^z\right)^2$ in terms of Ising variables. The resulting quadratic terms are then linearized by Hubbard-Stratonovich transformations to give\cite{Nishimori96,Bapst12}
\begin{equation}
Z_M=
\left(\sqrt{\frac{\beta J N}{\pi M}}\right)^{M}
\prod_{\kappa=0}^{M-1}
\int_{-\infty}^{\infty} dm(\kappa)
\exp
\left(
-\frac{\beta J N}{M}\sum_{\kappa=0}^{M-1}[m(\kappa)]^2
+
N\ln\mathcal{T}
\right)
,
\label{eq.sec_II.02}
\end{equation}
where 
\begin{equation}
\mathcal{T}
=
\sum_{\sigma=\pm1}
\langle \sigma |
\left[
\prod_{\kappa=0}^{M-1}
e^{\frac{1}{M} [ \beta \Gamma \sigma^x + 2\beta Jm(\kappa) \sigma^z  ]  }
\right]
| \sigma \rangle
\label{eq.sec_II.03}
\end{equation}
is the trace of a single spin with effective Hamiltonian 
\begin{equation}
\mathcal{H}^{\mathrm{eff}}(\kappa)
=
\beta\Gamma\sigma^x
+
2 \beta J m(\kappa) \sigma^z
,
\label{eq.sec_II.04}
\end{equation}
and $m(\kappa)$ is the order parameter (magnetization) introduced by the linearization at the $\kappa$th Trotter slice, $\sigma$ is an Ising variable taking values $\pm1$, and $|\sigma\rangle$ is the eigenvector of $\sigma^z$ with eigenvalue $\sigma$. In the limit $M\rightarrow\infty$, $Z$ takes the form of a path integral where one sums over all possible paths $m(\kappa)$. 

We first establish the relation between $\mathcal{T}$ and the differential equation Eq. (\ref{eq.sec_I.06}). To advance $|\psi(t)\rangle$ by an infinitesimal time step $\Delta t$, we have
\begin{eqnarray}
|\psi(t+\Delta t)\rangle
&=&
\left[
1
+
\mathcal{H}^{\mathrm{eff}}(t)
\Delta t
\right]
|\psi(t)\rangle
+
O[(\Delta t)^2]
\nonumber \\
&\approx&
e^{
\mathcal{H}^{\mathrm{eff}}(t)
\Delta t
}
|\psi(t)\rangle
.
\label{eq.sec_II.05}
\end{eqnarray}
The solution of $|\psi(t)\rangle$ at a later time $t+M\Delta t$ is obtained by consecutive application of Eq. (\ref{eq.sec_II.05}),
\begin{equation}
|\psi(t+M\Delta t)\rangle
=
\left[
\prod_{\kappa=0}^{M-1}
e^{\mathcal{H}^{\mathrm{eff}}(t+\kappa \Delta t)\Delta t}
\right]
|\psi(t)\rangle
.
\label{eq.sec_II.06}
\end{equation}
The product $\prod_{\kappa} e^{\mathcal{H}^{\mathrm{eff}}(t+\kappa \Delta t)\Delta t} $ is known as the fundamental matrix of Eq. (\ref{eq.sec_I.06}). In $\mathcal{T}$, it propagates the initial condition $|\sigma(0)\rangle=|\sigma\rangle$ from $t=0$ to $t=1$,
\begin{equation}
\mathcal{T}
=
\sum_{\sigma=\pm1}
\langle
\sigma(0)
|
\sigma(1)
\rangle.
\label{eq.sec_II.07}
\end{equation}
To calculate $\mathcal{T}$, we therefore need $|\sigma(t)\rangle$ the solution of the eigenvectors of $\sigma^z$ at time $t$. Since $\mathcal{H}^{\mathrm{eff}}(t)=\mathcal{H}_s+\lambda \mathcal{H}_d(t)$ with $\mathcal{H}_s=\beta \Gamma \sigma^x + 2 \beta J m_s \sigma^z$ and $\mathcal{H}_d(t)=2\beta J m_d(t) \sigma^z$, time-dependent perturbation theory\cite{Ballentine98} can be used to obtain a perturbative expansion of $|\sigma(t)\rangle$,
\begin{equation}
|\sigma(t)\rangle
=
|\sigma^{(0)}(t)\rangle
+
\lambda
|\sigma^{(1)}(t)\rangle
+
\lambda^2
|\sigma^{(2)}(t)\rangle
+
\cdots
,
\label{eq.sec_II.08}
\end{equation}
where $|\sigma^{(r)}(t)\rangle$ is the $r$th-order approximation of $|\sigma(t)\rangle$. Inserting Eq. (\ref{eq.sec_II.08}) into (\ref{eq.sec_II.07}), we have
\begin{equation}
\mathcal{T} = \mathcal{T}^{(0)} + \lambda\mathcal{T}^{(1)} + \lambda^2\mathcal{T}^{(2)} + \cdots,
\label{eq.sec_II.09}
\end{equation}
where 
\begin{equation}
\mathcal{T}^{(r)}
=
\sum_{\sigma=\pm1}
\langle \sigma(0) | \sigma^{(r)}(1) \rangle.
\label{eq.sec_II.10}
\end{equation}
The derivations of $|\sigma^{(r)}(t)\rangle$ and $\mathcal{T}^{(r)}$ are given in Appendix \ref{app.A.TDPT}. We now truncate $\mathcal{T}$ at the fourth order,
\begin{equation}
\mathcal{T}
\stackrel{\mathrm{4th}}{\longrightarrow}
\mathcal{T}^{(0)}
+
\lambda
\mathcal{T}^{(1)}
+
\lambda^2
\mathcal{T}^{(2)}
+
\lambda^3
\mathcal{T}^{(3)}
+
\lambda^4
\mathcal{T}^{(4)}
,
\label{eq.sec_II.11}
\end{equation}
where $\stackrel{\mathrm{4th}}{\longrightarrow}$ denotes `fourth-order approximation' and 
\begin{align}
\mathcal{T}^{(0)} &= 2\cosh\varepsilon \label{eq.sec_II.12}\\
\mathcal{T}^{(1)} &= (2\beta J)2\alpha\sinh\varepsilon \,M_0   \label{eq.sec_II.13}\\ 
\mathcal{T}^{(2)} &=
(2\beta J)^2
\{
2\alpha^2\cosh\varepsilon \, M_{00}
+
\gamma^2
[
e^{\varepsilon}
M_{-+}
+
e^{-\varepsilon}
M_{+-}
]
\}
\label{eq.sec_II.14}\\ 
\mathcal{T}^{(3)} &=
(2\beta J)^3
\{
2\alpha^3\sinh\varepsilon\,M_{000} 
\nonumber\\ 
&
+ \alpha\gamma^2 [ e^{\varepsilon}(M_{0-+}-M_{-0+} + M_{-+0})
 -e^{-\varepsilon} ( M_{0+-}-M_{+0-}  + M_{+-0})
]
\}
\label{eq.sec_II.15}\\ 
\mathcal{T}^{(4)} &=
(2\beta J)^4
\{
2\alpha^4\cosh\varepsilon\,M_{0000} 
+ 
\gamma^4[e^{\varepsilon}M_{-+-+} + e^{-\varepsilon}M_{+-+-}]
\nonumber\\ 
&
+
\alpha^2\gamma^2
[
e^{\varepsilon}(M_{-+00}-M_{-0+0}+M_{0-+0}-M_{0-0+}+M_{00-+}+M_{-00+})
\nonumber\\ 
&
+e^{-\varepsilon}(M_{+-00}-M_{+0-0}+M_{0+-0}-M_{0+0-}+M_{00+-}+M_{+00-})
]\}
\label{eq.sec_II.16}
\end{align}
where $\varepsilon=\sqrt{(\beta \Gamma)^2+(2\beta Jm_s)^2}$, $\alpha=\frac{2\beta J m_s}{\varepsilon}$, and $\gamma=-\frac{\beta\Gamma}{\varepsilon}$. We have also introduced the notation
\begin{equation} 
M_{s_1 \cdots s_k}
=
\int_0^1 dt_1
\,
m_d(t_{1}) e^{s_{1}2\varepsilon t_{1}}
\cdots
\int_0^{t_{k-1}} dt_k
\,
m_d(t_{k}) e^{s_{k}2\varepsilon t_{k}}
,
\label{eq.sec_II.17}
\end{equation} 
where $s_a$ ($a\in\{1,\cdots,k\}$) is the sign of the exponent of $e^{s_a 2\varepsilon t_a}$ and is either $+$, or $0$, or $-$. Inserting Eq. (\ref{eq.sec_II.11}) into Eq. (\ref{eq.sec_II.02}), $Z$ becomes
\begin{align}
Z
\stackrel{\mathrm{4th}}{\longrightarrow}\,
&
e^{-\beta N f_s} \int \mathcal{D}m_d(t) 
\times
\left[
1
+
\frac{1}{N}
\left(
V_4
+
\frac{1}{2}
(V_3)^2
\right)
\right]
\times
\nonumber\\
&
\exp 
\left[
-\beta J
\int_0^1dt \, m_d^2(t)
+
\frac{(2\beta J)^2 \gamma^2}{2\cosh\varepsilon}(e^{\varepsilon}M_{-+}+e^{-\varepsilon}M_{+-})
\right]
,
\label{eq.sec_II.18}
\end{align}
where $f_s$ is the static free energy per spin given by Eq. (\ref{eq.app_B.02}) and   
\begin{align}
V_3
=
&
L_3
-
L_1L_2
+
\frac{1}{3}\left(L_1\right)^3
,
\label{eq.sec_II.19}\\
V_4
=
&
L_4
-
L_1L_3
-
\frac{1}{2}
\left(L_2\right)^2
+
\left(L_1\right)^2L_2
-
\frac{1}{4}\left(L_1\right)^4
,
\label{eq.sec_II.20}
\end{align}
and $L_i=\frac{\mathcal{T}^{(i)}}{\mathcal{T}^{(0)}}$.


\section{The $N^0$ and $N^{-1}$ terms of $E_0$ }
\label{sec.III.N0 and N-1 terms of E0}

To evaluate Eq. (\ref{eq.sec_II.18}), expand $m_d(t)$ in Fourier series
\begin{equation}
m_d(t)
=
\sum_{n=-\infty}^{\infty}
c_n
\,
e^{i 2\pi n t}.
\label{eq.sec_III.01}
\end{equation}
Eq. (\ref{eq.sec_II.18}) becomes gaussian
\begin{equation}
Z
\stackrel{\mathrm{4th}}{\longrightarrow}
C
\,
e^{-\beta N f_s}
\int
dc_0
\prod_{n=1}^{\infty}
dc_n dc_n^*
\,
\left[
1
+
\frac{1}{N}
\left(
V_4
+
\frac{1}{2}
\left(
V_3
\right)^2
\right)
\right]
\,
\exp
\left(
-\beta J \sum_{n=-\infty}^{\infty}g_n c_n c_{-n}
\right)
,
\label{eq.sec_III.02}
\end{equation}
where $g_n=1-\frac{g}{(2\pi n)^2+(2\varepsilon)^2}$, $g=\frac{8\Gamma^2 J \beta^3\tanh\varepsilon}{\varepsilon}$, $dc_ndc_n^*=d\mathrm{Re}(c_n)d\mathrm{Im}(c_n)$, and $C=\sqrt{\frac{\beta J}{\pi}}\prod_{n=1}^{\infty}(\frac{2\beta J}{\pi})$. Performing the gaussian integrals and inserting  $Z$ into Eq. (\ref{eq.sec_I.02}), we obtain
\begin{equation}
E_0
\stackrel{\mathrm{4th}}{\longrightarrow}
Nf_s
+
\frac{\sqrt{\varepsilon^2 -\frac{g}{4} }-\varepsilon}{\beta}
-
\frac{1}{N}
\left(
\frac{z_4^1 + z_4^2 + z_3^1 +z_3^2}{\beta}
\right)
.
\label{eq.sec_III.03}
\end{equation}
We first discuss the $N^0$ term. Inserting $m_s$ given by Eq. (\ref{eq.app_B.03}), we have
\begin{equation}
N^0 \,\mathrm{term}
=
\left\{
\begin{array}{ccc}
 \sqrt{\Gamma(\Gamma-2J)}-\Gamma  & \mathrm{for} & \Gamma \ge2J, \\
 \sqrt{(2J)^2-\Gamma^2} -2J  & \mathrm{for} & \Gamma <2J. \\
\end{array}
\right.
\label{eq.sec_III.04}
\end{equation}
In Fig. \ref{fig.fig01}(a), Eq. (\ref{eq.sec_III.04}) is plotted and compared with the results of numerical calculations.

\begin{figure}[h]
\begin{center}
\includegraphics[scale=1.0]{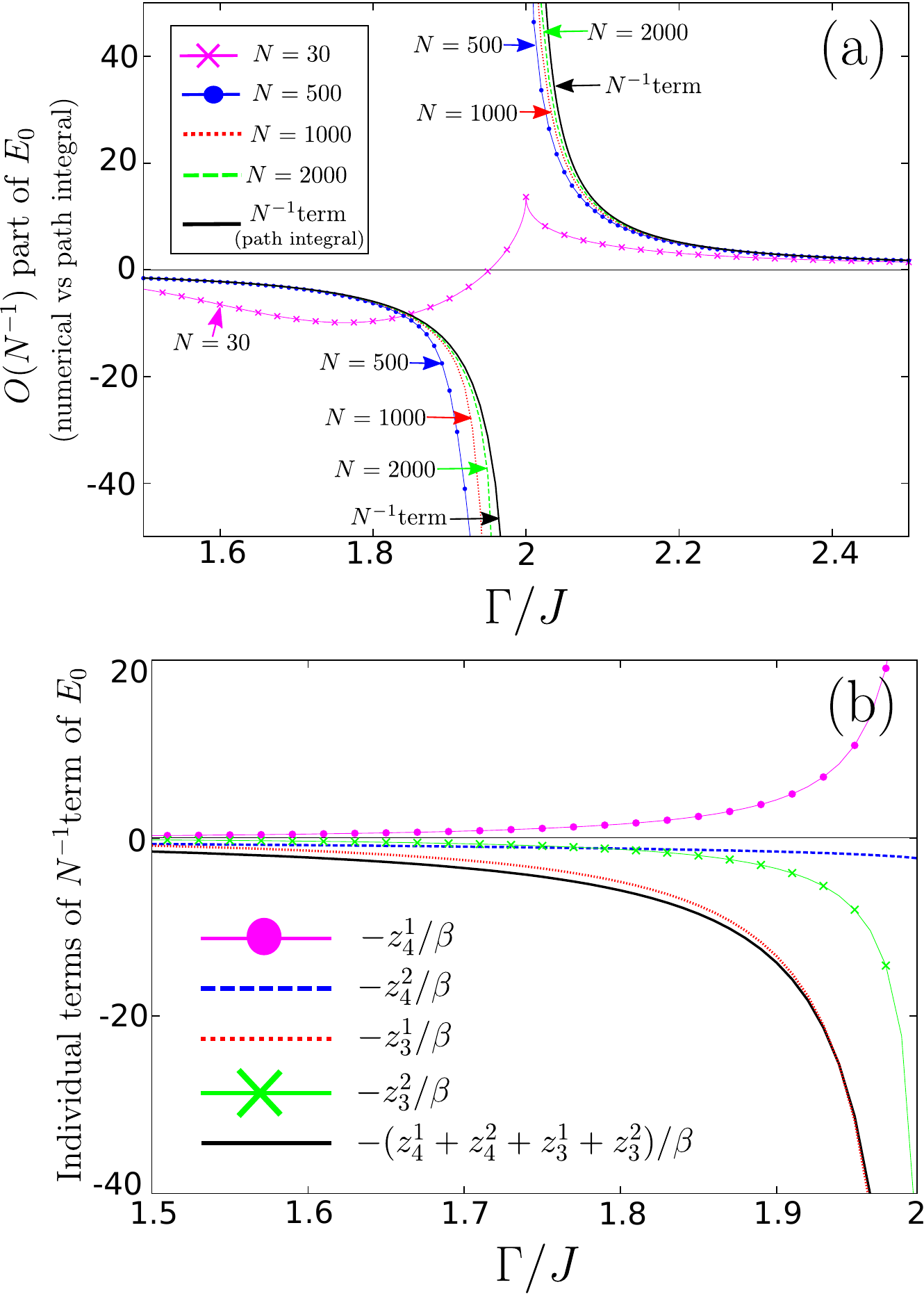}
\caption{(a) The $O(N^{-1})$ part of $E_0$ of the ferromagnetic model. For the curves of various $N$, the free energy and $N^0$ term are subtracted from the numerically computed $E_0$ and the result multiplied by $N$. The $N^{-1}$ term (black solid line) is the second correction to the free energy obtained by incorporating dynamical paths into the path integral. (b) Individual terms of the $N^{-1}$ term in the ferromagnetic phase.}
\label{fig.fig02}
\end{center}
\end{figure}

For the $N^{-1}$ term, $z_4^1$ and $z_4^2$ originate from integrating over $V_4$ and the other two terms from $\frac{1}{2}(V_3)^2$. These terms and their derivations are given in Appendix \ref{app.C.derivation z412z312}. We first consider the paramagnetic phase. In this phase, only $z_4^1$ is non-zero. From Eq. (\ref{eq.app_C.06}), we have
\begin{equation}
N^{-1}\,\mathrm{term}
=
-\frac{z_4^1}{\beta}
=
\frac{J(2\Gamma-J)}{2(\Gamma-2J)}
-
\frac{\Gamma J}{\sqrt{\Gamma(\Gamma-2J)}}
\,\,\,\,\,\,
\mathrm{for}
\,\,
\Gamma \ge2J
.
\label{eq.sec_III.05}
\end{equation}
In Fig. \ref{fig.fig02}(a), we plotted Eq. (\ref{eq.sec_III.05}) (black solid line) in the paramagnetic phase. Results of numerical calculations are also shown where for each $N$ the free energy and $N^0$ term are subtracted away from $E_0$ and the result multiplied by $N$. It is seen that Eq. (\ref{eq.sec_III.05}) agrees with the numerical results. 

In the ferromagnetic phase, all four terms contribute. $z_4^1$ and $z_3^1$ are given by Eqs. (\ref{eq.app_C.08}) and (\ref{eq.app_C.09}) while $z_4^2$ and $z_3^2$ are evaluated numerically. In Fig. \ref{fig.fig02}(a), we plotted $-\frac{1}{\beta}(z_4^1+z_4^2+z_3^1+z_3^2)$ (black solid line) in the region $\Gamma<2J$. It is seen that the curve agrees with the results of numerical calculation. Fig. \ref{fig.fig02}(b) shows the individual terms that make up the $N^{-1}$ term in the ferromagnetic phase. 

Fig. \ref{fig.fig02}(a) shows that the $N^{-1}$ term diverges at the critical point. This can be understood by examining the rate at which the minimum point on the curves of $E_0-Nf_s$ in Fig. \ref{fig.fig01}(a) converge towards the critical value of $-2$ at $\Gamma=2J$ as $N$ increases. We found numerically that the difference between the critical and finite-$N$ value decreases as $N^{-0.33}$. Upon multiplying by $N$, this decrease is turned into an increase that scales as $N^{0.67}$, thereby accounting for the divergence.

\section{Calculation of $Z_Q$ and energy gap in the paramagnetic phase}
\label{sec.IV.ZQ}

In the Introduction we derived Eq. (\ref{eq.sec_I.10}) and proposed applying it to the paramagnetic phase of Eq. (\ref{eq.sec_I.03}) with $Q$ given by Eq. (\ref{eq.sec_I.12}). To see that this could work, consider the ground and first excited-states when $J$ is small. The former is 
\begin{equation}
|E_0\rangle_{J=0}
=\prod_{i=1}^{N}
|\sigma_i^x=+1\rangle,
\label{eq.sec_IV.01}
\end{equation}
where all spins point along the positive $x$-direction. From first-order perturbation theory, the latter is 
\begin{equation}
|E_1\rangle_{J=0}
=
\frac{1}{\sqrt{N}}
\sum_{i=1}^N |i\rangle
,
\label{eq.sec_IV.02}
\end{equation}
where $|i\rangle$ is the state where the $i$th spin in $|E_0\rangle_{J=0}$ is flipped. $|E_0\rangle_{J=0}$ and $|E_1\rangle_{J=0}$ are both non-degenerate and have parity eigenvalues $+1$ and $-1$, respectively. As $J$ increases beyond the perturbative regime, these eigenvalues must still remain the same because $Q$ is conserved. The conditions for using Eq. (\ref{eq.sec_I.10}) are therefore satisfied. 

The path integral of $Z_Q$ has the same form as $Z$ but with $\mathcal{T}$ replaced by  
\begin{equation}
\mathcal{T}_Q
=
\sum_{\sigma=\pm1}
\langle \sigma(0)|
\sigma^x
| \sigma(1) \rangle,
\label{eq.sec_IV.03}
\end{equation}
where one multiplies $|\sigma(1)\rangle$ by $\sigma^x$ before taking the inner product. We truncate $\mathcal{T}_Q$ at second-order, 
\begin{equation}
\mathcal{T}_Q
\stackrel{\mathrm{2nd}}{\longrightarrow}
\mathcal{T}_Q^{(0)}
+
\lambda
\mathcal{T}_Q^{(1)}
+
\lambda^2
\mathcal{T}_Q^{(2)}
,
\label{eq.sec_IV.04}
\end{equation}
where $\stackrel{\mathrm{2nd}}{\longrightarrow}$ denotes `second-order approximation' and 
\begin{align}
\mathcal{T}_Q^{(0)}
&
=
-2\gamma \sinh\varepsilon
\label{eq.sec_IV.05}\\
\mathcal{T}_Q^{(1)}
&
=
(2\beta J)
\,
\alpha\gamma
\left[
-2\cosh\varepsilon \,M_0
+
e^{\varepsilon}M_-
+
e^{-\varepsilon}
M_+
\right]
\label{eq.sec_IV.06}\\
\mathcal{T}_Q^{(2)}
&
=
(2\beta J)^2
\{
-2\alpha^2\gamma \sinh\varepsilon \, M_{00} - \gamma^3 (e^{\varepsilon}M_{-+}-e^{-\varepsilon}M_{+-})
\nonumber\\
&
+
\alpha^2\gamma
[
e^{\varepsilon}(M_{0-} - M_{-0})
+
e^{-\varepsilon}(M_{+0} - M_{0+})
]
\}
\label{eq.sec_IV.07}
\end{align}
are derived in the same manner as for $\mathcal{T}^{(r)}$. $Z_Q$ becomes
\begin{equation}
Z_Q
\stackrel{\mathrm{2nd}}{\longrightarrow}
e^{-\beta N \hat{f}_s}
\int \mathcal{D}m_d(t)
\exp
\left[
-\beta J
\int_0^1 dt \, m^2_d(t)
+
\frac{(2\beta J)^2\gamma^2}{2\sinh \varepsilon}
(
e^{\varepsilon} M_{-+}
-
e^{-\varepsilon} M_{+-}
)
\right]
,
\label{eq.sec_IV.08}
\end{equation}
where $\hat{f}_s=J m_s^2-\frac{1}{\beta}\ln(-2\gamma\sinh\varepsilon)$ and $m_s$ is given by Eq. (\ref{eq.app_B.03}) in the limit $\beta\rightarrow\infty$. In Eq. (\ref{eq.sec_IV.08}), we have substituted $m_s=0$ in the paramagnetic phase.

The path integral is performed by expanding $m_d(t)$ as
\begin{equation}
m_d(t)
=
\sideset{}{'}
\sum_{n}
c_n
\,
e^{i\pi n t}
,
\label{eq.sec_IV.09}
\end{equation}
where the boundary condition is $m_d(0)=-m_d(1)$ and $\sum_n^{'}$ sums over all positive and negative odd integers. Eq. (\ref{eq.sec_IV.08}) becomes
\begin{equation}
Z_Q
\stackrel{\mathrm{2nd}}{\longrightarrow}
\hat{C}
\,
e^{-\beta N \hat{f}_s}
\int
\sideset{}{'}
\prod_{n}
dc_n dc_n^*
\exp
\left(
-\beta J
\sideset{}{'}
\sum_n
\hat{g}_n
c_n c_{-n}
\right),
\label{eq.sec_IV.10}
\end{equation}
where $\hat{g}_n=1-\frac{\hat{g}}{(\pi n)^2+(2\varepsilon)^2}$, $\hat{g}=\frac{8\Gamma^2 J \beta^3\mathrm{coth}\varepsilon}{\varepsilon}$, $\prod_n^{'}$ runs over the positive odd integers, and $\hat{C}=\prod_n^{'}(\frac{2\beta J}{\pi})$. Performing the gaussian integrals, we get
\begin{equation}
Z_Q
\stackrel{\mathrm{2nd}}{\longrightarrow}
e^{-\beta N \hat{f}_s}
\frac{\cosh \varepsilon}{\cosh\sqrt{\varepsilon^2-\frac{\hat{g}}{4}}}
\,\,\,\,\,\,\,\,\,
\mathrm{for}
\,\,\,
\Gamma\ge 2J
.
\label{eq.sec_IV.11}
\end{equation}
Inserting Eq. (\ref{eq.sec_IV.11}) and
\begin{equation}
Z
\stackrel{\mathrm{2nd}}{\longrightarrow}
e^{-\beta N f_s}
\frac{\sinh \varepsilon}{\sinh\sqrt{\varepsilon^2-\frac{g}{4}}}
,
\label{eq.sec_IV.12}
\end{equation}
into the relation Eq. (\ref{eq.sec_I.10}), we get 
\begin{equation}
E_1
\stackrel{\mathrm{2nd}}{\longrightarrow}
-N\Gamma -\Gamma + 3\sqrt{\Gamma(\Gamma-2J)}
\,\,\,\,\,\,\,\,\,\,
\mathrm{for} 
\,\,\,\,
\Gamma\ge 2J.
\label{eq.sec_IV.13}
\end{equation}
Subtracting away the second-order approximation of $E_0$, we get
\begin{equation}
E_1-E_0
\stackrel{\mathrm{2nd}}{\longrightarrow}
2\sqrt{\Gamma(\Gamma-2J)}
\,\,\,\,\,\,\,\,\,\,
\mathrm{for} 
\,\,\,\,
\Gamma\ge 2J.
\label{eq.sec_IV.14}
\end{equation}
In Fig. \ref{fig.fig01}(b), Eq. (\ref{eq.sec_IV.14}) is plotted and compared with the results of numerical calculations.

In deriving Eq. (\ref{eq.sec_I.10}), we have required that the ground and first excited-states be of opposite parity. From Fig. \ref{fig.fig01}(b), however, we see that $E_0$ and $E_1$ collapse together in the ferromagnetic phase, and the ground-state becomes doubly-degenerate and mixed in parity. In particular, when $\Gamma$ is small the ground-state is spanned by the two states
\begin{equation}
|E_0^{\pm}\rangle_{\Gamma=0}
=
\frac{1}{\sqrt{2}}
\left(
\prod_{i=1}^N
|\sigma_i^z=+1\rangle
\pm
\prod_{i=1}^N
|\sigma_i^z=-1\rangle
\right),
\label{eq.sec_IV.15}
\end{equation}
where the superscript $\pm$ labels the parity eigenvalues. From second-order perturbation theory, the first excited-state is also doubly-degenerate and mixed in parity,
\begin{equation}
|E_1^{\pm}\rangle_{\Gamma=0}
=
\frac{1}{\sqrt{N}}
\sum_{i=1}^N	
|i\rangle_{\mp}
,
\label{eq.sec_IV.16}
\end{equation}
where $|i\rangle_{\pm}$ is the state where the $i$th spin in $|E_0^{\pm}\rangle_{\Gamma=0}$ is flipped. The subtraction of Eqs. (\ref{eq.sec_I.08}) and (\ref{eq.sec_I.09}) is therefore no longer effective, and the relation Eq. (\ref{eq.sec_I.10}) is not applicable in the ferromagnetic phase.

\section{Energy gap in both phases}
\label{sec.V.gap in both phases}

\subsection{Trace formula for $E_0+E_1$}
\label{subsec.VA.E0+E1}

\begin{figure}[h]
\begin{center}
\includegraphics[scale=0.7]{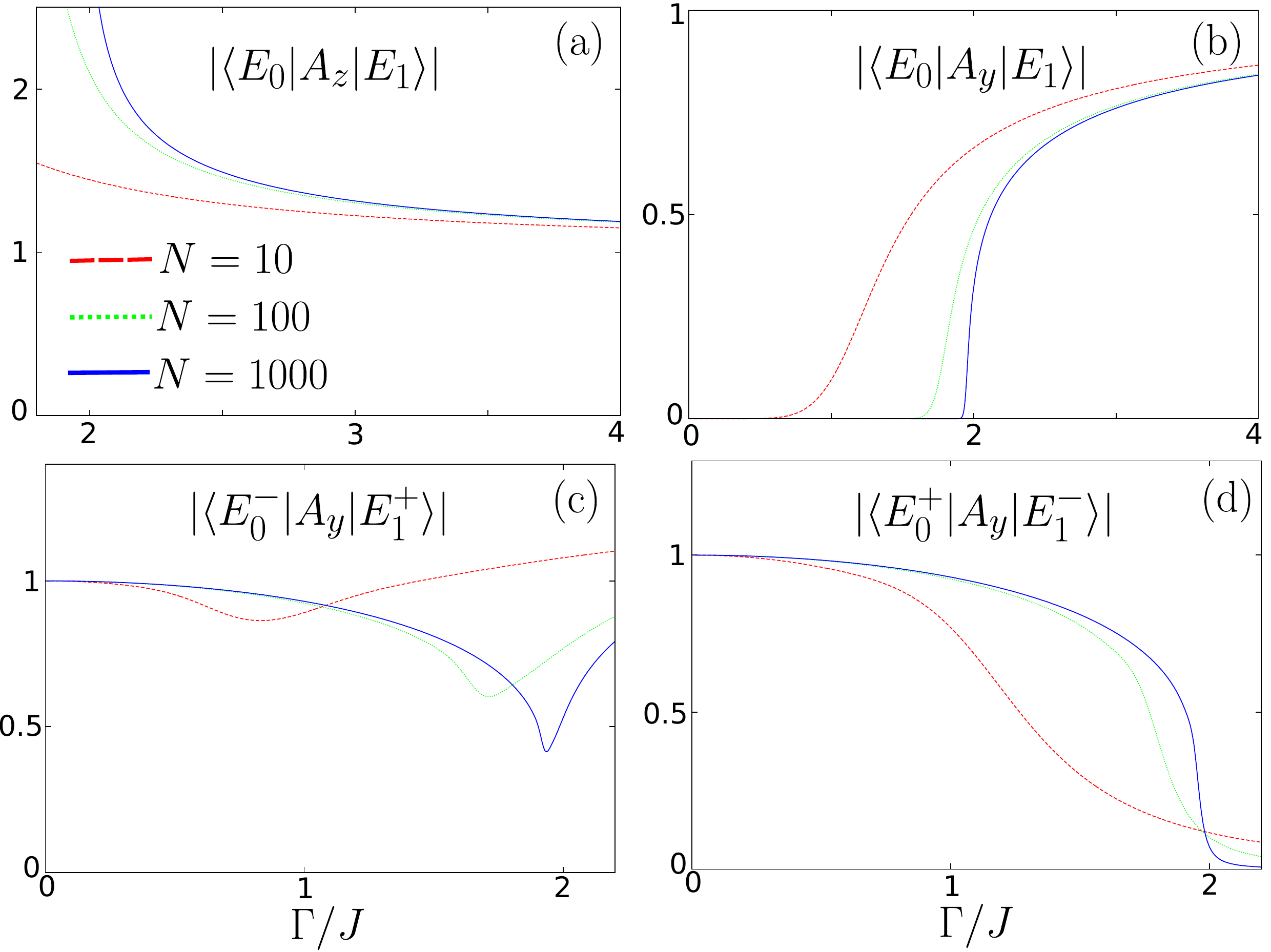}
\caption{Graphs showing the non-vanishing of $\sum_{a,b}|\langle E_0^a|A_{\mu}|E_1^b\rangle|^2$ for the ferromagnetic model. The matrix elements are computed numerically and their absolute values plotted for various $N$. (a) $|\langle E_0|A_z|E_1\rangle|$ in the paramagnetic phase. (b) $|\langle E_0|A_y|E_1\rangle|$. In the ferromagnetic phase ($\Gamma<2J$), the matrix element vanishes because it becomes $\langle E_0^+|A_y|E_0^-\rangle$. (c) $|\langle E_0^-|A_y|E_1^+\rangle|$ in the ferromagnetic phase.  (d) $|\langle E_0^+|A_y|E_1^-\rangle|$ in the ferromagnetic phase.}
\label{fig.fig03}
\end{center}
\end{figure}

Let us define the operator
\begin{equation}
A_{\mu}
=
\frac{1}{\sqrt{N}}
\sum_{i=1}^N
\sigma^{\mu}_i
.
\label{eq.sec_VA.01}
\end{equation}
Consider 
\begin{align}
e^{-\beta H}A_ye^{-\beta H}
&
=
e^{-2\beta E_0} \left( \sum_{a,b} |E_0^a\rangle\langle E_0^a|A_y|E_0^b\rangle\langle E_0^b| \right)
+
e^{-\beta (E_0+E_1)} \left(\sum_{a,b}|E_0^a\rangle\langle E_0^a|A_y|E_1^b\rangle\langle E_1^b| \right)
\nonumber\\
&
+
e^{-\beta (E_0+E_1)}\left( \sum_{a,b}|E_1^a\rangle\langle E_1^a|A_y|E_0^b\rangle\langle E_0^b| \right)
+
e^{-2\beta E_1}\left( \sum_{a,b}|E_1^a\rangle\langle E_1^a|A_y|E_1^b\rangle\langle E_1^b| \right)
+
\cdots,
\label{eq.sec_VA.02}
\end{align}
where $a$ in $|E_n^a\rangle$ denotes the parity of the energy state. Now, $\langle E_0^a|A_y|E_0^b\rangle=0$. This is because $[H,A_z]=2i\Gamma A_y$, so $2i\Gamma\langle E_0^{a}|A_y|E_0^{b}\rangle=\langle E_0^{a}|[H,A_z]|E_0^{b}\rangle=(E_0-E_0)\langle E_0^{a}|A_z|E_0^{b}\rangle=0$. Hence $e^{-2\beta E_0}$ vanishes and $e^{-\beta(E_0+E_1)}$ becomes the leading term. Squaring Eq. (\ref{eq.sec_VA.02}) and then taking trace, we have
\begin{equation}
\mathrm{Tr}
\left(
A_ye^{-2\beta H}A_ye^{-2\beta H}
\right)
=
2
e^{-2\beta(E_0+E_1)}
\left(
\sum_{a,b}
|\langle E_0^a|A_y| E_1^b\rangle|^2
\right)
\left[
1
+
O\left(e^{-\beta(E_2-E_1)}\right)
\right].
\label{eq.sec_VA.03}
\end{equation}
Assuming that $\sum_{a,b}|\langle E_0^a|A_y| E_1^b\rangle|^2$ does not vanish, we have
\begin{equation}
E_0+E_1
=
\lim_{\beta\rightarrow\infty}
-
\frac{1}{2\beta}
\ln Z_{A_y}
,
\label{eq.sec_VA.04}
\end{equation}
where
\begin{equation}
Z_{A_y}
=
\mathrm{Tr}
\left(
A_ye^{-2\beta H}A_ye^{-2\beta H}
\right)
.
\label{eq.sec_VA.05}
\end{equation}
To check that $\sum_{a,b}|\langle E_0^a|A_y| E_1^b\rangle|^2$ does not vanish, we computed the matrix elements numerically. In Fig. \ref{fig.fig03}(b) we plotted $|\langle E_0|A_y|E_1\rangle|$ in the paramagnetic phase for various $N$ and in Figs. \ref{fig.fig03}(c) and (d) we plotted $|\langle E_0^- |A_y|E_1^+\rangle|$ and $|\langle E_0^+ |A_y|E_1^-\rangle|$ in the ferromagnetic phase. It is seen that the matrix elements are non-zero and so Eq. (\ref{eq.sec_VA.04}) is valid in both phases. 

Eq. (\ref{eq.sec_VA.04}) is also valid in the paramagnetic phase with $A_y$ replaced by $A_z$. In this case, the matrix element $\langle E_0|A_z|E_0\rangle$ vanishes because $QA_z=-A_zQ$, so $\langle E_0^+|A_z|E_0^+ \rangle=\langle E_0^+ |QQA_zQQ|E_0^+ \rangle=\langle E_0^+ |QA_zQ|E_0^+ \rangle=-\langle E_0^+ |A_zQQ|E_0^+ \rangle=-\langle E_0^+ |A_z|E_0^+ \rangle=0$. Numerical calculation of $|\langle E_0|A_z|E_1\rangle|$ shown in Fig. \ref{fig.fig03}(a) also confirms that it is non-zero in the paramagnetic phase.

\subsection{Leading approximation of the path integral of $Z_{A_{\mu}}$}
\label{subsec.VB.leading approx of ZAmu}

Inserting Eq. (\ref{eq.sec_VA.01}) into Eq. (\ref{eq.sec_VA.05}), and all spins being identical in Eq. (\ref{eq.sec_I.03}), we have
\begin{equation}
Z_{A_{\mu}}
=
\mathrm{Tr}
\left(
\sigma_i^{\mu}
e^{-2\beta H}
\sigma_i^{\mu}
e^{-2\beta H}
\right)
+
(N-1)
\mathrm{Tr}
\left(
\sigma_i^{\mu}
e^{-2\beta H}
\sigma_j^{\mu}
e^{-2\beta H}
\right)
.
\label{eq.sec_VB.01}
\end{equation}
When deriving the path integral, we need to factor out the spins with indices $i$ or $j$. From the Suzuki-Trotter decomposition of $Z_{A_{\mu}}$ we have
\begin{equation}
\left(
Z_{A_{\mu}}
\right)_M
=
\left(
\frac{2\beta J N}{\pi M}
\right)^{M}
\prod_{\kappa=0}^{2M-1}
\int_{-\infty}^{\infty}
dm(\kappa)
\frac{\mathcal{T}_0\mathcal{T}_{3\mu} + (N-1) \mathcal{T}_{1\mu}\mathcal{T}_{2\mu}}{(\mathcal{T}_0)^2}
\exp
\left(
-\frac{2\beta JN}{M}\sum_{\kappa=0}^{2M-1}[m(\kappa)]^2
+
N
\ln
\mathcal{T}_0
\right)
,
\label{eq.sec_VB.02}
\end{equation}
where
\begin{align}
\mathcal{T}_{0}
&
=
\sum_{\sigma=\pm1}
\langle \sigma |
\left[
\prod_{\kappa=0}^{2M-1}
e^{\frac{1}{M}[2\beta\Gamma\sigma^x+4\beta J m(\kappa)\sigma^z]}
\right]
|\sigma\rangle
.
\label{eq.sec_VB.03}\\[10pt]
\mathcal{T}_{1\mu}
&
=
\sum_{\sigma=\pm1}
\langle \sigma |
\sigma^{\mu}
\left[
\prod_{\kappa=0}^{2M-1}
e^{\frac{1}{M}[2\beta\Gamma\sigma^x+4\beta J m(\kappa)\sigma^z]}
\right]
|\sigma\rangle
.
\label{eq.sec_VB.04}\\[10pt]
\mathcal{T}_{2\mu}
&
=
\sum_{\sigma=\pm1}
\langle \sigma |
\left[
\prod_{\kappa=M}^{2M-1}
e^{\frac{1}{M}[2\beta\Gamma\sigma^x+4\beta J m(\kappa)\sigma^z]}
\right]
\sigma^{\mu}
\left[
\prod_{\kappa=0}^{M-1}
e^{\frac{1}{M}[2\beta\Gamma\sigma^x+4\beta J m(\kappa)\sigma^z]}
\right]
|\sigma\rangle
.
\label{eq.sec_VB.05}\\[10pt]
\mathcal{T}_{3\mu}
&
=
\sum_{\sigma=\pm1}
\langle \sigma |
\sigma^{\mu}
\left[
\prod_{\kappa=M}^{2M-1}
e^{\frac{1}{M}[2\beta\Gamma\sigma^x+4\beta J m(\kappa)\sigma^z]}
\right]
\sigma^{\mu}
\left[
\prod_{\kappa=0}^{M-1}
e^{\frac{1}{M}[2\beta\Gamma\sigma^x+4\beta J m(\kappa)\sigma^z]}
\right]
|\sigma\rangle
.
\label{eq.sec_VB.06}
\end{align}
The leading approximation of $Z_{A_{\mu}}$ is obtained by expanding the $\mathcal{T}_0$ inside the exponent to second order and all the other single spin traces to zeroth or first order. For $\mathcal{T}_{2\mu}$ and $\mathcal{T}_{3\mu}$, as the time evolution is interrupted halfway by a Pauli matrix, let us introduce the notation
\begin{equation}
M_s^{t_1,t_2}
=
\int_{t_1}^{t_2}
dt
\,
m_d(t)
\,
e^{s4\varepsilon t}
,
\label{eq.sec_VB.07}
\end{equation}
where $s$ has the same meaning as in Eq. (\ref{eq.sec_II.17}). For $\mu=y$, we have
\begin{align}
\mathcal{T}_{1y}
&
\stackrel{\mathrm{1st}}{\longrightarrow}
i \lambda (4\beta J)  \gamma
(
e^{4\varepsilon} M_{-}^{0,2} 
-
e^{-4\varepsilon} M_{+}^{0,2} 
)
\label{eq.sec_VB.08}\\
\mathcal{T}_{2y}
&
\stackrel{\mathrm{1st}}{\longrightarrow}
i \lambda (4\beta J) \gamma
(
M_-^{0,1}
-
M_+^{0,1}
+
e^{8\varepsilon}
M_-^{1,2}
-
e^{-8\varepsilon}
M_+^{1,2}
)
\label{eq.sec_VB.09}\\
\mathcal{T}_{3y}
&
\stackrel{\mathrm{s.a.}}{\longrightarrow}
2
\label{eq.sec_VB.10}
\end{align}
where $\stackrel{\mathrm{1st}}{\longrightarrow}$ and $\stackrel{\mathrm{s.a.}}{\longrightarrow}$ denote `first-order' and `static' approximation, respectively. For $\mu=z$, we have
\begin{align}
\mathcal{T}_{1z}
&
\stackrel{\mathrm{1st}}{\longrightarrow}
2\alpha\sinh 4\varepsilon
+
\lambda
(4\beta J)
[
2
\alpha^2
\cosh 4\varepsilon
\,
M_0^{0,2}
+
\gamma^2
(
e^{4\varepsilon}
M_-^{0,2}
+
e^{-4\varepsilon}
M_+^{0,2}
)
]
\label{eq.sec_VB.11}\\
\mathcal{T}_{2z}
&
\stackrel{\mathrm{1st}}{\longrightarrow}
2\alpha\sinh4\varepsilon
+
\lambda(4\beta J)
[
2\alpha^2\cosh 4\varepsilon
\,
M_0^{0,2}
+
\gamma^2
(
M_+^{0,1}
+
M_-^{0,1}
+
e^{8\varepsilon}
M_-^{1,2}
+
e^{-8\varepsilon}
M_+^{1,2}
)
]
\label{eq.sec_VB.12}\\
\mathcal{T}_{3z}
&
\stackrel{\mathrm{s.a.}}{\longrightarrow}
2
(
\alpha^2 \cosh 4\varepsilon
+
\gamma^2
)
\label{eq.sec_VB.13}
\end{align}

\subsection{Calculation of $Z_{A_y}$ and the energy gap }
\label{subsec.VC.calculate ZAy}

The path integral Eq. (\ref{eq.sec_VB.02}) is performed by expanding 
\begin{equation}
m_d(t)
=
\sum_{n=-\infty}^{\infty}
c_n
\,
e^{i\pi n t}
,
\label{eq.sec_VC.01}
\end{equation}
where $0<t<2$. For $\mu=y$, Eq. (\ref{eq.sec_VB.02}) becomes
\begin{align}
Z_{A_{y}}
\stackrel{\mathrm{l.a.}}{\longrightarrow}
&
\,
C'
e^{-\beta N f'_s}
\int
dc_0
\prod_{n=1}^{\infty}
dc_n dc_n^*
\,
\exp
\left(
-
4\beta J
\sum_{n=-\infty}^{\infty}
g'_n c_n c_{-n}
\right)
\nonumber\\
&
\times
\left[
\mathrm{sech}4\varepsilon
-
\left(\frac{g'}{4\beta\Gamma}\right)^2
\sum_{n=-\infty}^{\infty}
c_n c_{-n}
\frac{(-1)^n (\pi n)^2 }{[(\pi n)^2+(4\varepsilon)^2]^2}
\right]
,
\label{eq.sec_VC.02}
\end{align}
where $f'_s=4Jm_s^2-\frac{1}{\beta}\ln 2 \cosh 4\varepsilon$, $g'_n=1-\frac{g'}{(\pi n)^2+(4\varepsilon)^2}$, $g'=\frac{32\Gamma^2 J \beta^3 \tanh4\varepsilon}{\varepsilon}$, $C'=\sqrt{\frac{4\beta J}{\pi}}\prod_{n=1}^{\infty}(\frac{8\beta J}{\pi})$, and $\stackrel{\mathrm{l.a.}}{\longrightarrow}$ denotes `leading approximation'. Performing the gaussian integrals, we obtain 
\begin{equation}
Z_{A_y}
\stackrel{\mathrm{l.a.}}{\longrightarrow}
e^{-\beta N f'_s}
\,
\frac{\sinh4\varepsilon \tanh4\varepsilon}{\sinh^2 4\sqrt{\varepsilon^2-\frac{g'}{16}}}
\,
\frac{\sqrt{\varepsilon^2-\frac{g'}{16}}}{\varepsilon}
.
\label{eq.sec_VC.03}
\end{equation}
The term $\sqrt{\varepsilon^2-\frac{g'}{16}}$ is finite and non-zero everywhere except at the critical point where it vanishes. Inserting Eq. (\ref{eq.sec_VC.03}) into Eq. (\ref{eq.sec_VA.04}), we get 
\begin{equation}
E_0+E_1
\stackrel{\mathrm{l.a.}}{\longrightarrow}
\frac{N}{2}
f'_s
+
4
\sqrt{
\Gamma^2
+
(2Jm_s)^2
-
\frac{2\Gamma^2 J}{\sqrt{\Gamma^2 + (2Jm_s)^2}}
}
-2
\sqrt{\Gamma^2 + (2Jm_s)^2}
.
\label{eq.sec_VC.04}
\end{equation}
Inserting $m_s$ from Eq. (\ref{eq.app_B.03}) into Eq. (\ref{eq.sec_VC.04}) and subtracting $2E_0$, we obtain 
\begin{equation}
E_1-E_0
\stackrel{\mathrm{l.a.}}{\longrightarrow}
\left\{
\begin{array}{ccc}
2\sqrt{\Gamma(\Gamma-2J)}  & \mathrm{for} & \Gamma \ge2J, \\
2\sqrt{(2J)^2-\Gamma^2}    & \mathrm{for} & \Gamma <2J. \\
\end{array}
\right.
\label{eq.sec_VC.05}
\end{equation}
In Fig. \ref{fig.fig01}(b), Eq. (\ref{eq.sec_VC.05}) is plotted and compared with the results of numerical calculations. For completeness, the calculation of $Z_{A_z}$ is also given in Appendix \ref{app.D.Calculation of ZAz}.

%

\section{Applications to the Hopfield model}
\label{sec.2nd_reply_PRB.Hopfield_Model}

In the previous sections, we have demonstrated and verified our approach on a simple model. In this section, we apply it to a non-trivial disordered mean-field model, the Hopfield model in a transverse field. The Hamiltonian is
\begin{equation}
H^{\mathrm{HM}}
=
-
\frac{1}{2N}
\sum_{\mu=1}^p
\left(
\sum_{i=1}^N
\xi_i^{\mu}
\sigma_i^z
\right)^2
-
\Gamma \sum_{i=1}^{N}\sigma_i^x,
\label{eq.Hopfield_section.01}
\end{equation}
where the random variables $\xi^{\mu}_i$ are each $+1$ or $-1$ with equal probability. The original Hopfield model, without the transverse field, was proposed within the context of associative memory where $p$ $N$-dimensional binary vectors $\vec{\xi}^{\mu}$ ($\mu=1,\cdots,p$) are first `memorized' and later retrieved by the temporal evolution of the system\cite{Hopfield82,Amit89}. It was later generalized to the form Eq. (\ref{eq.Hopfield_section.01}) by various authors\cite{Ma93a,Ma93b,Nishimori96}. Unlike the ferromagnetic model, $H^{\mathrm{HM}}$ does not commute with the total angular momentum and so must be diagonalized in the full Hilbert space whose dimension scales exponentially with $N$. This makes it difficult to study the exact numerical properties of the ground-state energy and excitation gap for large system sizes. On the other hand, the thermodynamics of the model has been studied in detail by Nishimori and Nonomura using path integral with static approximation at both low and high memory loadings\cite{Nishimori96}. Here, we focus on the low loading regime where $p$ is kept fixed and $N$ is scaled up. In this regime, one does not need to average over the disorder using the replica trick. We can therefore study a system which is realized by a specific set of patterns and so anomalies peculiar to specific realizations are not masked by any averaging process.

Nishimori and Nonomura's paper shows that at $\Gamma=1$ the system undergoes a second-order transition from a paramagnetic to a condensed phase where multiple minima form on the free energy surface. One group of minima consists of the so-called odd-mixtures. These states have multiple macroscopic overlaps (i.e. magnetizations) with an odd number of patterns and are local minima on the energy surface. For our purpose, we shall focus on another group of states called the memory states. These have overlap with only one pattern and are the ground-states in the entire condensed phase. According to the authors' analysis, the set of $p$ memory states appear spontaneously at $\Gamma=1$ and are all degenerate. The magnetization and free energy of these states are similar to that of the ferromagnetic model and are given by substituting $J\rightarrow\frac{1}{2}$ into Eqs. (\ref{eq.app_B.03}) and (\ref{eq.app_B.04}), respectively. A crucial step when deriving these quantities involves replacing the average over spin sites [c.f. Eq. (\ref{eq.Hopfield_section.02}) below] by an average over the disorder. Although exact in the thermodynamic limit, this neglects certain small but significant disorder-induced effects in finite-sized systems. As an example, consider when $\Gamma=0$. From Eq. (\ref{eq.app_B.04}) the free energy is $-\frac{N}{2}$; from Eq. (\ref{eq.Hopfield_section.01}), however, one sees that the exact ground-state energy contains contributions from overlaps of the retrieved pattern with the other patterns. The magnitudes of these overlaps are of order $O(1)$ and cannot be neglected when incorporating quantum fluctuations and computing energy gaps. This need for careful treatment of the disorder when analyzing energy gaps was emphasized recently by Knysh\cite{Knysh16}.

In the following subsection, we first revisit the static approximation of the partition function of the Hopfield model, taking care to incorporate small random effects due to disorder. In sections \ref{subsec.2nd_reply_PRB.dynamical fluctuations on E0 of Hopfield} and \ref{subsec.2nd_reply_PRB.para gap using ZQ of Hopfield}, we apply the techniques developed earlier to improve upon the free energy as well as calculate various energy gaps of the system. In section \ref{subsec.2nd_reply_PRB.anomalous transitions in Hopfield}, we briefly discuss the occurrence of `anomalous' transitions encountered in disordered systems.

\subsection{Static approximation incorporating the effects of disorder}
\label{subsec.2nd_reply_PRB.fs_w_disorder}

The derivation of the static free energy per spin is similar to that of the ferromagnetic model and is given in Ref.\cite{Nishimori96}. In the limit $\beta\rightarrow\infty$, it is
\begin{equation}
f_s^{\mathrm{HM}}
=
\frac{1}{2N}
\sum_{\mu=1}^p
(m_s^{\mu})^2
-
\frac{1}{N}
\sum_{i=1}^N
\sqrt{\Gamma^2 
+
\frac{1}{N}
\left(
\sum_{\mu=1}^p
\xi_i^{\mu}
m_s^{\mu}
\right)^2
}
,
\label{eq.Hopfield_section.02}
\end{equation}
where $m_s^{\mu}$ ($\mu=1,\cdots,p$) are static approximations of the order parameters $m^{\mu}(t)$ introduced to linearize the quadratic terms $(\sum_i \xi_i^{\mu}\sigma_i^z)^2$. These `magnetizations' signify the overlap between the spins and the $\mu$th pattern. Note that we have rescaled the order parameters in Eq. (\ref{eq.Hopfield_section.02}) to account for both macroscopic and non-macroscopic overlaps. A macroscopic overlap has magnitude $\propto\sqrt{N}$ in Eq. (\ref{eq.Hopfield_section.02}). Traditionally, the site average $\frac{1}{N}\sum_i$ on the right hand side of Eq. (\ref{eq.Hopfield_section.02}) is replaced by an average over disorder $\frac{1}{2^{p}}\sum_{\xi^{\mu}}$ at a single site. We shall keep the site average in our analysis. 

\begin{figure}[h]
\begin{center}
\includegraphics[scale=1.0]{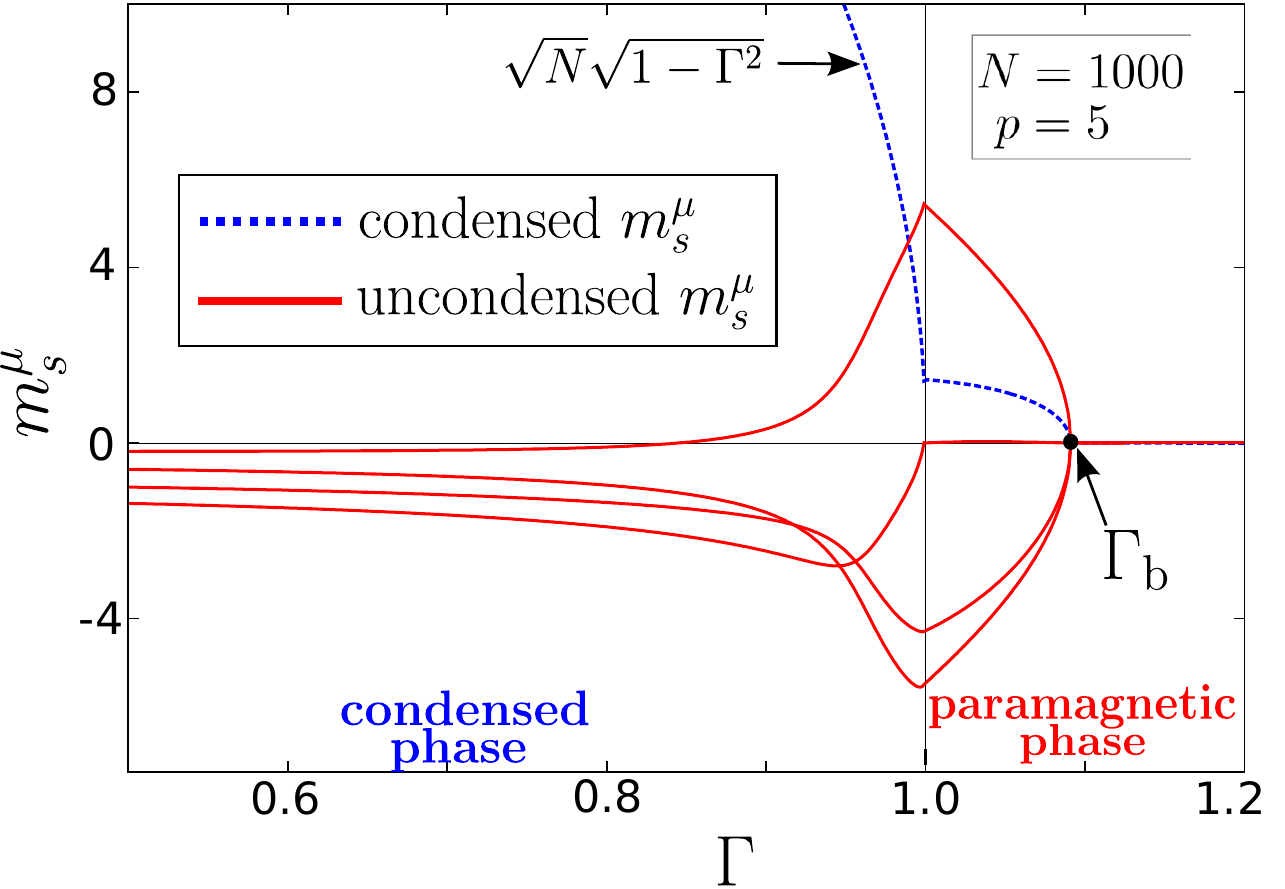}
\caption{Numerical solutions of Eqs. (\ref{eq.Hopfield_section.03}) for a particular realization of $p=5$ patterns ($N=1000$). Solid (red) lines show the variables $m_s^{\mu}$ that are uncondensed in both phases. The dashed (blue) line shows the one that magnetizes macroscopically upon entering the condensed phase. At $\Gamma_{\mathrm{b}}$ the zero solution becomes unstable and non-zero solutions appear.}
\label{fig.fig04}
\end{center}
\end{figure}

The $m_s^{\mu}$ are obtained by solving the set of $p$ equations $\partial f_s^{\mathrm{HM}}/\partial m_s^a=0$, or
\begin{equation}
m_s^a
=
\frac{1}{N}
\sum_{i=1}^N
\frac
{
\left(
\sum_{\mu=1}^p
\xi_i^a\xi_i^{\mu}m_s^{\mu}
\right)
}
{
\sqrt{\Gamma^2 + \frac{1}{N} 
\left(
\sum_{\mu=1}^p
\xi_i^{\mu}m_s^{\mu}
\right)^2}
}
\label{eq.Hopfield_section.03}
,
\end{equation}
for $a=1,\cdots,p$. Our method of solving Eqs. (\ref{eq.Hopfield_section.03}) is where we depart from the traditional treatment. Previously, only condensed variables (i.e. with macroscopic magnetization) acquire non-zero values; uncondensed variables are assumed to be zero. Here, we allow non-zero solutions of Eq. (\ref{eq.Hopfield_section.03}) even for the uncondensed variables. In the paramagnetic phase, we optimize over all $p$ variables $m_s^{\mu}$ to find the minimum $f_s^{\mathrm{HM}}$. In the condensed phase, for the memory states, one of the $m_s^{\mu}$ is fixed at the macroscopic value $\sqrt{N}\sqrt{1-\Gamma^2}$ while $f_s^{\mathrm{HM}}$ is minimized over the remaining $p-1$ uncondensed variables. We solve Eq. (\ref{eq.Hopfield_section.03}) this way numerically and the results are shown in Fig. \ref{fig.fig04} for a particular realization of $p=5$ patterns with $N=1000$. The $m_s^{\mu}$ that are uncondensed in both phases are plotted using solid (red) lines, while the $m^{\mu}_s$ that magnetizes macroscopically upon entering the condensed phase is plotted using dashed (blue) line. Note that we have shown only one out of 5 possible results of the condensed phase. In the paramagnetic phase, as $\Gamma$ decreases $(m^1_s,\cdots,m^p_s)=(\cdots,0,\cdots)$ is the only solution until $\Gamma_{\mathrm{b}}$ where the magnetizations suddenly acquire non-zero values. This is because the zero solution becomes unstable due to the smallest eigenvalue of the Hessian matrix
\begin{equation}
\left.
\frac{\partial^2f_s^{\mathrm{HM}}}{\partial m_s^{\mu}\partial m_s^{\nu}}
\right|_{(\cdots,0,\cdots)}
=
(N\Gamma)^{-1}
\left[
(\Gamma-1)\delta^{\mu\nu}
-
O^{\mu\nu}
\right]
\label{eq.Hopfield_section.04}
\end{equation}
turning negative. $\delta^{\mu\nu}$ is the Kronecker delta and $O^{\mu\nu}=\frac{1}{N}\sum_{i=1}^N\xi^{\mu}_i\xi_i^{\nu}$ ($O^{\mu\mu}=0$) is the inter-pattern overlap matrix. Hence, $\Gamma_{\mathrm{b}}$ is given by $1+\omega^{\mathrm{max}}$ where $\omega^{\mathrm{max}}$ is the largest eigenvalue of the overlap matrix. It is different for different realizations of patterns. At exactly $\Gamma=1$ there is an ambiguity in the condensed variable as its macroscopic magnetization is supposed to be zero. Numerically, however, it is observed that for large $N$ the overall continuity of the solutions is maintained if we skip the critical point and resume calculations at a $\Gamma$ slightly below it. At $\Gamma=0$, we obtain for the uncondensed variables $m_s^{\mu}=\sqrt{N}O^{a\mu}$ where $m_s^a$ is the condensed variable, recovering the classical result.

\begin{figure}[h]
\begin{center}
\includegraphics[scale=0.9]{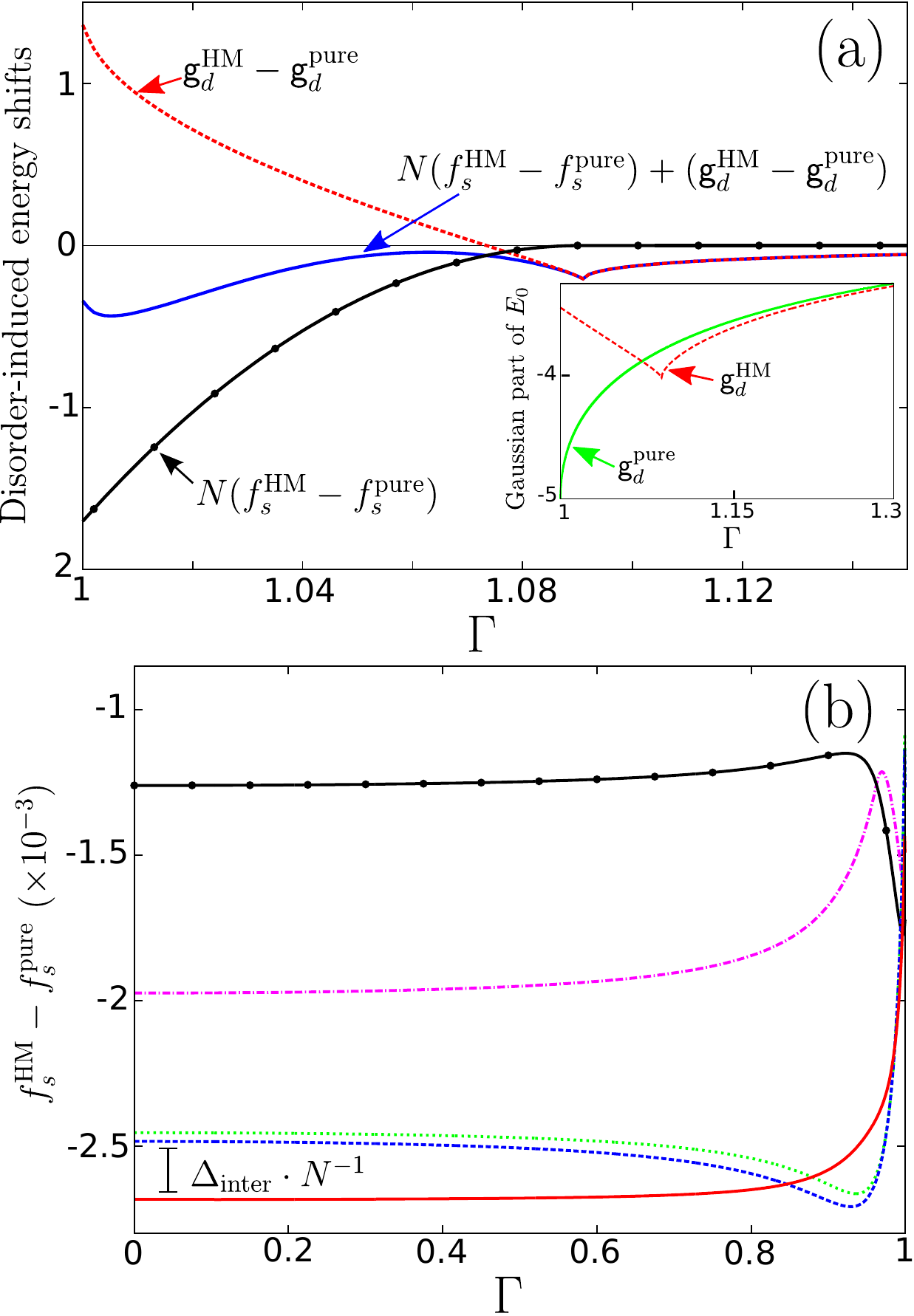}
\caption{Comparing various energies of the Hopfield model with that of a pure (i.e., ferromagnetic) system. Results for the system with the magnetization curves of Fig. \ref{fig.fig04} are shown. (a) Paramagnetic phase. Solid (black) line with circles shows $N(f_s^{\mathrm{HM}}-f_s^{\mathrm{pure}})$. Dashed (red) line shows the gaussian correction $\mathsf{g}_d^{\mathrm{HM}}-\mathsf{g}_d^{\mathrm{pure}}$. Solid (blue) line shows the shift of the ground-state energy of the Hopfield model given by Eq. (\ref{eq.Hopfield_section.06}) from that of a pure system. Inset: Graphs of $\mathsf{g}_d^{\mathrm{HM}}$ (red dashed line) and $\mathsf{g}_d^{\mathrm{pure}}$ (green solid line). (b) Condensed phase. Graphs of $f_s^{\mathrm{HM}}-f_s^{\mathrm{pure}}$ for all 5 memory states. The memory state from Fig. \ref{fig.fig04} is plotted using solid (black) line with circles. The inter-pattern energy gap $\Delta_{\mathrm{inter}}$ is defined as the energy difference between the two memory states with the lowest energies, as shown.}
\label{fig.fig05}
\end{center}
\end{figure}

Fig. \ref{fig.fig05} shows the $f_s^{\mathrm{HM}}$ corresponding to the magnetization curves of Fig. \ref{fig.fig04}. To highlight the effects of disorder, let us refer to the ferromagnetic model as a pure system and denote the free energy given by Eq. (\ref{eq.app_B.04}) (with $J\rightarrow\frac{1}{2}$) as $Nf_s^{\mathrm{pure}}$. In Fig. \ref{fig.fig05}(a), the graph of $N(f_s^{\mathrm{HM}}-f_s^{\mathrm{pure}})$ in the paramagnetic phase is plotted using solid (black) line with circles. From scale of the vertical axis, we see that disorder changes the energy by an amount comparable to that caused by quantum fluctuations in a pure system [c.f. Fig. \ref{fig.fig01}(a)]. Fig. \ref{fig.fig05}(b) shows the graphs of $f_s^{\mathrm{HM}}-f_s^{\mathrm{pure}}$ for all 5 memory states in the condensed phase. The graph corresponding to the magnetization curves of Fig. \ref{fig.fig04} is plotted in solid (black) line with circles. By taking into account the effects of disorder, we see that the memory states are actually not degenerate but different in energies.

\subsection{Effects of dynamical paths on the ground-state energy}
\label{subsec.2nd_reply_PRB.dynamical fluctuations on E0 of Hopfield}

\subsubsection{Incorporation of gaussian fluctuations}
\label{subsubsec.2nd_reply_PRB.gaussian fluctuations for E0 of Hopfield}

We now apply the method of Secs. \ref{sec.II.expansions of T and Z} and \ref{sec.III.N0 and N-1 terms of E0} to calculate the contributions of dynamical paths to the ground-state energy. For simplicity, we shall consider the leading gaussian fluctuations [i.e. expanding to second order in Eq. (\ref{eq.sec_II.09})]. The derivation of Sec. \ref{sec.II.expansions of T and Z} remains the same and one simply replaces $J\rightarrow\frac{1}{2}$, $m_s\rightarrow \frac{1}{\sqrt{N}}\sum_{\mu=1}^p \xi_i^{\mu}m_s^{\mu}$, $\varepsilon\rightarrow\varepsilon_i=\beta\sqrt{\Gamma^2+\frac{1}{N}(\sum_{\mu}\xi_i^{\mu}m_s^{\mu})^2}$, and $m_d(t)\rightarrow\sum_{\mu}\xi_i^{\mu}m_d^{\mu}(t)$ in Eqs. (\ref{eq.sec_II.12}) to (\ref{eq.sec_II.14}). The main difference from the ferromagnetic model comes from the site average of Eq. (\ref{eq.sec_II.14})
\begin{equation}
\frac{1}{N}
\sum_{i=1}^N
\frac
{
e^{\varepsilon_i}M_{-+}
+
e^{-\varepsilon_i}M_{+-}
}
{
2\varepsilon_i^2\cosh\varepsilon_i
}
\approx
\sum_{n=-\infty}^{\infty}
\frac{
\frac{2\tanh\tilde{\varepsilon}}{\tilde{\varepsilon}}
}{(2\pi n)^2+(2\tilde	{\varepsilon})^2}
\left(
\sum_{\mu=1}^p
c^{\mu}_n c^{\mu}_{-n}
+
\sum_{\mu\ne\nu}
O^{\mu\nu}
c_n^{\mu}c_{-n}^{\nu}
\right)
,
\label{eq.Hopfield_section.05}
\end{equation}
where $c^{\mu}_n$ is expansion coefficient in the Fourier expansion of $m_d^{\mu}(t)$ [c.f. Eq. (\ref{eq.sec_III.01})]. One encounters an additional term on the right hand side of Eq. (\ref{eq.Hopfield_section.05}) where disorder introduces coupling between the paths $m_d^{\mu}(t)$ via the inter-pattern overlap matrix $O^{\mu\nu}$. The $\approx$ in Eq. (\ref{eq.Hopfield_section.05}) stems from the approximation $\varepsilon_i\approx \tilde{\varepsilon}=\beta\sqrt{\Gamma^2+ \frac{1}{N} \sum_{\mu=1}^p(m_s^{\mu})^2}$ which introduces an error that is smaller by a factor of $N^{-\frac{1}{2}}$ compared to the leading term and can be neglected for large $N$. The terms inside the parenthesis of Eq. (\ref{eq.Hopfield_section.05}) give a $p$-dimensional gaussian integral that can be readily integrated. The approximate ground-state energy of the Hopfield model including gaussian fluctuations is then
\begin{equation}
E_0
\approx
Nf_s^{\mathrm{HM}}
+
\frac{1}{\beta}
\left(
\sum_{\mu=1}^p
\sqrt{
\tilde{\varepsilon}^2
-
\frac{\beta^3\Gamma^2(1+\omega^\mu)}{\tilde{\varepsilon}}
}
-
p\tilde{\varepsilon}
\right)
,
\label{eq.Hopfield_section.06}
\end{equation}
where $\omega^{\mu}$ is the $\mu$th eigenvalue of the overlap matrix. Let us denote the gaussian contribution to $E_0$ [the second term on the right hand side of Eq. (\ref{eq.Hopfield_section.06})] by $\mathsf{g}_d^{\mathrm{HM}}$. The corresponding term of a pure system, given by $p$ times the $N^{0}$ term in Eq. (\ref{eq.sec_III.03}), is denoted by $\mathsf{g}_d^{\mathrm{pure}}$. The inset of Fig. \ref{fig.fig05}(a) shows $\mathsf{g}_d^{\mathrm{HM}}$ (red dashed line) in the paramagnetic phase for the system with the magnetization curves of Fig. \ref{fig.fig04}. The graph of $\mathsf{g}_d^{\mathrm{pure}}$ (green solid line) is also shown for comparison. The difference $\mathsf{g}_d^{\mathrm{HM}}-\mathsf{g}_d^{\mathrm{pure}}$ is the gaussian contribution to $E_0$ induced by disorder, and is shown in the main plot using dashed (red) line. In Fig. \ref{fig.fig05}(a), the solid (blue) line shows the shift of $E_0$ given by Eq. (\ref{eq.Hopfield_section.06}) from that of a pure system. The observation that $\mathsf{g}_d^{\mathrm{HM}}-\mathsf{g}_d^{\mathrm{pure}}$ and $N(f^{\mathrm{HM}}_s-f^{\mathrm{pure}}_s)$ partially cancel each other is specific to the system shown and is not generalizable to other realizations of patterns.

\subsubsection{Inter-pattern energy gap in the condensed phase}
\label{subsubsec.2nd_reply_PRB.interpattern energy gap of Hopfield}

\begin{figure}[h]
\begin{center}
\includegraphics[scale=0.9]{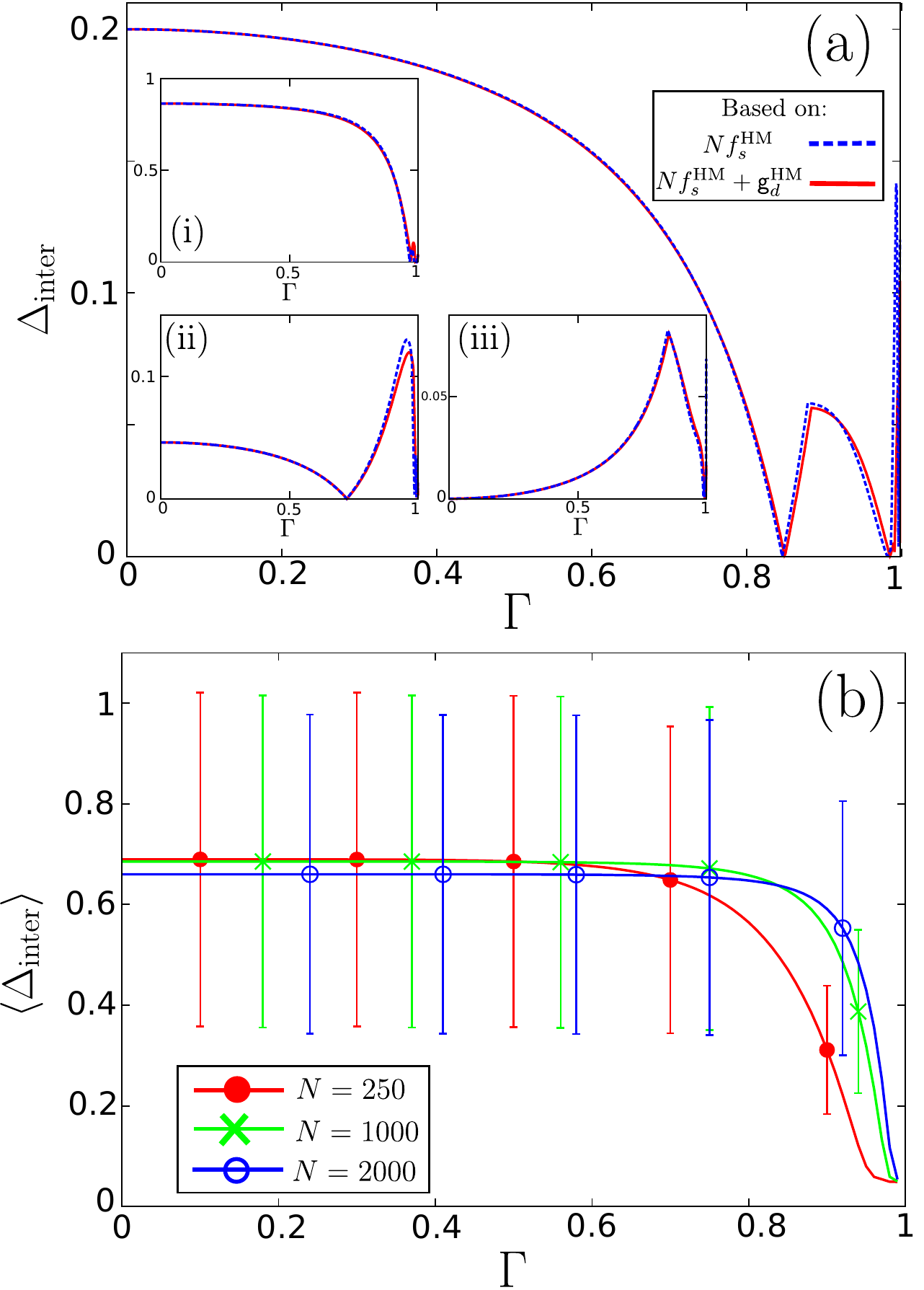}
\caption{Inter-pattern energy gap $\Delta_{\mathrm{inter}}$ of the Hopfield model in the condensed phase. (a) Graphs of $\Delta_{\mathrm{inter}}$ of particular realizations of patterns with $p=5$ and $N=1000$. Main plot shows the $\Delta_{\mathrm{inter}}$ obtained from the realization of Fig. \ref{fig.fig05}(b). Dashed (blue) line shows the $\Delta_{\mathrm{inter}}$ when the energy of the memory states is approximated by $Nf_s^{\mathrm{HM}}$; solid (red) line shows the case when the energy is approximated by $Nf_s^{\mathrm{HM}}+\mathsf{g}_d^{\mathrm{HM}}$. Insets (i)-(iii): Other examples of $\Delta_{\mathrm{inter}}$ with different realizations of patterns (same $p$ and $N$). (b) Graphs of the mean gap, $\langle\Delta_{\mathrm{inter}}\rangle$, for various $N$ ($p=5$). Each particular $\Delta_{\mathrm{inter}}$ is calculated by approximating the energy of the memory states by $Nf_s^{\mathrm{HM}}$. The curve of each $N$ is obtained by averaging over 5000 different realizations. Error bars indicate the standard deviation associated with the mean.}
\label{fig.fig06}
\end{center}
\end{figure}

The energy of each of the memory states in the condensed phase is also given by Eq. (\ref{eq.Hopfield_section.06}). We now examine one aspect of the gaussian fluctuations. Let us define the inter-pattern energy gap $\Delta_{\mathrm{inter}}$ as the energy difference between the two memory states with the lowest energies. If the energy of the memory states is approximated by $Nf_s^{\mathrm{HM}}$, then $\Delta_{\mathrm{inter}}$ is as shown in Fig. \ref{fig.fig05}(b). For clarity, this $\Delta_{\mathrm{inter}}$ is redrawn in the main plot of Fig. \ref{fig.fig06}(a) using dashed (blue) line. We now calculate the energy of the memory states again using Eq. (\ref{eq.Hopfield_section.06}), recalculate $\Delta_{\mathrm{inter}}$, and the result is plotted using solid (red) line. It is seen that the effects of gaussian fluctuations on $\Delta_{\mathrm{inter}}$ is quite small. This observation can be generalized to other realizations of patterns when $N$ is large. Insets (i)-(iii) of Fig. \ref{fig.fig06}(a) show other examples of $\Delta_{\mathrm{inter}}$ with different realizations of patterns (same $p$ and $N$). Hence, at least for the Hopfield model studied here, the main contribution to $\Delta_{\mathrm{inter}}$ comes from the disorder in $f_s^{\mathrm{HM}}$; the effects of dynamical fluctuations are quite minimal. 

Fig. \ref{fig.fig06}(b) shows the mean gap, $\langle \Delta_{\mathrm{inter}} \rangle$, obtained by averaging over different realizations of patterns, for various $N$ ($p=5$). The $\Delta_{\mathrm{inter}}$ of each realization is calculated by approximating the energy of each memory state by $Nf_s^{\mathrm{HM}}$. For each $N$, the average is taken over 5000 realizations of patterns. The error bars indicate the standard deviation associated with the mean. The results show that the mean gap is constant throughout most of the condensed phase. Unlike the free energy, $\Delta_{\mathrm{inter}}$ is not a self-averaging quantity. This can be discerned from the large standard deviations whose magnitudes remain fairly constant as $N$ increases. Another way to see this is to notice in Fig. \ref{fig.fig06}(a) the seemingly random and different forms of $\Delta_{\mathrm{inter}}$ exhibited by each particular realization of patterns, even though the system size of $N=1000$ is already quite large. One should therefore not draw conclusions about the $\Delta_{\mathrm{inter}}$ of specific systems based on the average $\langle \Delta_{\mathrm{inter}} \rangle$.

\subsection{Excitation gap in the paramagnetic phase}
\label{subsec.2nd_reply_PRB.para gap using ZQ of Hopfield}

We now consider the excitation gap in the paramagnetic phase using the method of $Z_Q$ presented in Sec. \ref{sec.IV.ZQ}. The Hamiltonian $H^{\mathrm{HM}}$ commutes with $Q$ given by Eq. (\ref{eq.sec_I.12}), so the ground and first excited-states are parity eigenstates and our derivation in the Introduction should be valid. One caveat is that even though the parity of the first excited-state is conserved, it might still change if the energy level encounters degeneracies (`collisions') with some higher levels before reaching the critical point. For the ferromagnetic model, it can be checked numerically that this does not happen. For $H^{\mathrm{HM}}$, first-order perturbation theory shows that near $\Gamma=\infty$ the first excited-state is non-degenerate and has parity $-1$. However, it is not possible to numerically obtain the exact first excited-state until the critical point for large system sizes. Here, we assume that for all realizations of $H^{\mathrm{HM}}$ the first excited-state has parity $-1$ in the entire paramagnetic phase.

\begin{figure}[h]
\begin{center}
\includegraphics[scale=1.0]{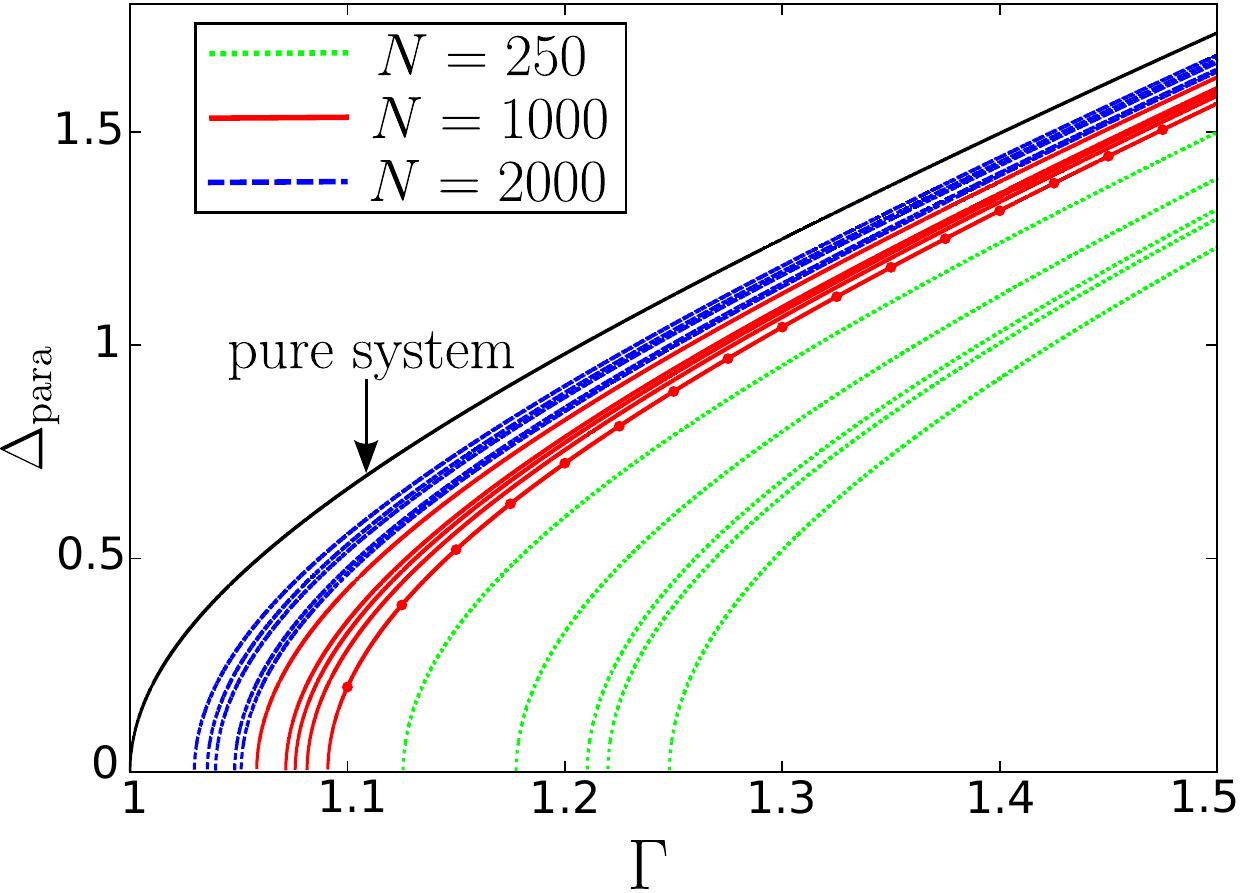}
\caption{Results for $\Delta_{\mathrm{para}}$, the excitation gap of the Hopfield model in the paramagnetic phase, approximated by Eq. (\ref{eq.Hopfield_section.07}), for various $N$ ($p=5$). Each curve represents the gap of one particular realization of patterns. Five different realizations are plotted for each $N$. The gap of a pure system is also shown for comparison. The solid (red) curve with circles corresponds to the system with the magnetization curves of Fig. \ref{fig.fig04}.} 
\label{fig.fig07}
\end{center}
\end{figure}

The calculation of $Z_Q$ for the Hopfield model is similar to that of $Z$ in the previous subsection. The result is analogous to Eqs. (\ref{eq.sec_IV.11}) and (\ref{eq.sec_IV.12}) where the `sinh' appearing in $Z$ are replaced by `cosh' in $Z_Q$. Denoting the excitation gap in the paramagnetic phase by $\Delta_{\mathrm{para}}$, as $\beta\rightarrow\infty$, we obtain
\begin{equation}
\Delta_{\mathrm{para}}
\approx
2
\sqrt{\Gamma[\Gamma-(1+\omega^{\mathrm{max}})]}
\,\,.
\label{eq.Hopfield_section.07}
\end{equation}
The $\approx$ in Eq. (\ref{eq.Hopfield_section.07}) stems from the approximation made in Eq. (\ref{eq.Hopfield_section.05}) and the neglect of terms beyond the gaussian fluctuations. Fig. \ref{fig.fig07} shows $\Delta_{\mathrm{para}}$ for various $N$ ($p=5$). To convey a sense of the effects of disorder, for each $N$ the gaps of five different realizations of patterns are shown. The gap of a pure system (i.e., $\omega^{\mathrm{max}}=0$) is also plotted for comparison. The approximate $\Delta_{\mathrm{para}}$ closes at $\Gamma_{\mathrm{b}}$. For finite systems, this closure is lifted by going beyond the gaussian fluctuations and should be understood as a point of crossover. Unlike $\Delta_{\mathrm{inter}}$, the gap $\Delta_{\mathrm{para}}$ is self-averaging, since $\omega^{\mathrm{max}}\rightarrow 0$ as $N\rightarrow\infty$. Below $\Gamma_{\mathrm{b}}$, the symmetry of the paramagnetic phase (i.e. the zero solution) is broken, giving rise to degeneracy in the ground and first-excited states. Eq. (\ref{eq.sec_I.10}) is no longer valid and the gap cannot be obtained below $\Gamma_{\mathrm{b}}$ via the current approach.


\subsection{Rough energy landscape and anomalous transitions in disordered systems}
\label{subsec.2nd_reply_PRB.anomalous transitions in Hopfield}

In pure systems, the transitions between different phases are relatively simple, most being either second or first-order in nature. In disordered models, however, the system may undergo other forms of transitions even within a particular thermodynamic phase. In a recent detailed study of the two-pattern Gaussian Hopfield model, Knysh showed that the presence of disorder inevitably leads to an effective system diffusing on a rough and random potential energy landscape\cite{Knysh16}. Depending on the realization of patterns, this potential might exhibit competing energy minima even within the condensed phase, leading to the need to consider `anomalous' (i.e. neither second nor first-order) transitions on top of the traditional macroscopic ones. Such transitions lead to bottlenecks in the gap structure of the glassy phase and have important implications for quantum annealing.

\begin{figure}[h]
\begin{center}
\includegraphics[scale=1.0]{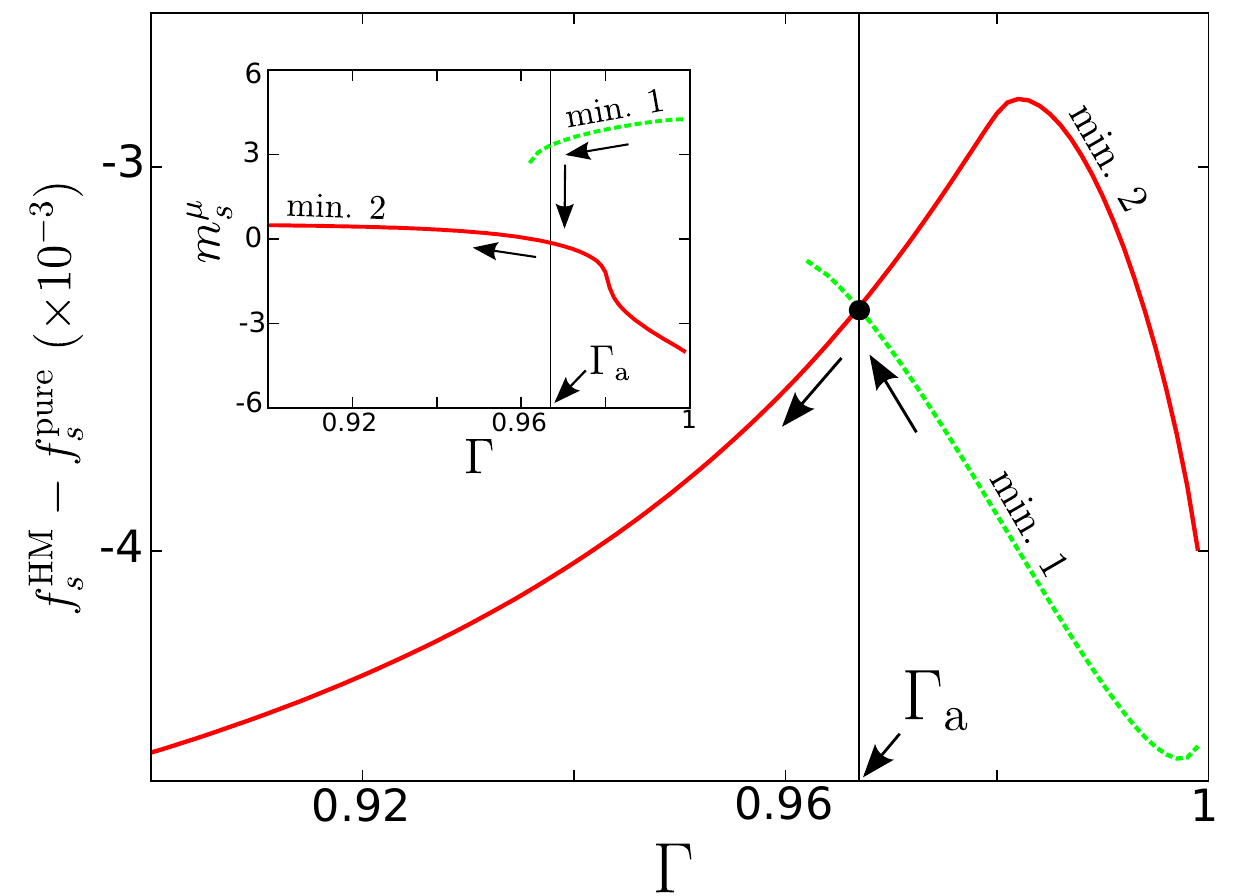}
\caption{Example of an anomalous transition occuring within the basin a particular memory state ($p=5$ and $N=250$). The main plot shows that two minima exist on the energy surface of $f_s^{\mathrm{HM}}$, labelled `min. 1' (green dashed line) and `min. 2' (red solid line). As $\Gamma$ decreases from 1, min. 1 is the ground-state. At $\Gamma_{\mathrm{a}}$ an anomalous transition occurs, after which min. 2 becomes the ground-state. Inset: Evolution of one of the magnetizations $m_s^{\mu}$ as $\Gamma$ decreases. At $\Gamma_{\mathrm{a}}$, the magnetization change discontinuously as it jumps from min.1 to min. 2.} 
\label{fig.fig08}
\end{center}
\end{figure}

In Sec. \ref{subsubsec.2nd_reply_PRB.interpattern energy gap of Hopfield}, we have seen that in the Hopfield model resonance can happen between the ground-states of two different memory states. The energy barrier separating the basins of memory states is of $O(N)$, so the closure of the gap $\Delta_{\mathrm{inter}}$ should be first-order like. On the other hand, the anomalous transitions described by Knysh take place within the basin of a macroscopic phase and the energy barriers are much smaller. One might ask if such transitions also occur for the Hopfield model we studied, and the answer is positive. Fig. \ref{fig.fig08} shows an example of an anomalous transition occuring within the basin of a particular memory state for a realization of patterns with $p=5$ and $N=250$. Two minima, labelled `min. 1' (green dashed line) and `min. 2' (red solid line), exist on the energy surface of $f^{\mathrm{HM}}_s$. As $\Gamma$ decreases from 1, min. 1 is the ground-state. At $\Gamma_{\mathrm{a}}$, the two minima becomes equal in energy and there is an anomalous transition. Below $\Gamma_{\mathrm{a}}$, min. 2 becomes the ground-state. Min. 1 disappears from the energy surface slightly below $\Gamma_{\mathrm{a}}$. The transition is discontinuous in nature. The inset shows the evolution of $m_s^{\mu}$ (only one component is shown) as $\Gamma$ decreases. It is seen that at $\Gamma_{\mathrm{a}}$ the magnetization makes a discontinuous jump from min.1 to min. 2. 

In our study, anomalous transitions are observed within the basins of the memory states as well as in the paramagnetic phase, and tend to occur in the vicinity of the critical point $\Gamma=1$. Local minima usually disappear as one proceeds deeper into the condensed or paramagnetic phase. As we did not encounter many local minima within each basin (usually $\le 3$), we performed an exhaustive search through the $p$-dimensional space to find them all. Note, however, that our study focuses only on binary patterns and for a low $p$ (=5). For other types of patterns and for larger $p$ values, the landscape of $f_s^{\mathrm{HM}}$ should become more rugged and the anomalous transitions might occur over a wider range of $\Gamma$ and more frequently. This, for instance, has been discussed by Knysh for the Gaussian Hopfield model\cite{Knysh16}.

\section{Discussions and conclusion}
\label{sec.VI.conclusion}

In this paper, we went beyond the static approximation of mean-field quantum spin models by incorporating dynamical paths into the path integral. The time-dependence of the trace of the time-ordered exponential of the effective Hamiltonian is calculated by first mapping it onto a time-dependent ordinary differential equation and then using perturbation theory to obtain a perturbative expansion of the trace. We derived two formulae, Eqs. (\ref{eq.sec_I.10}) and (\ref{eq.sec_VA.04}), for calculating the gap. We applied our method to an ordered and a disordered model. For the infinite-range ferromagnetic Ising model in a transverse field, we calculated the $N^0$ and $N^{-1}$ terms of $E_0$ and the energy gap in both phases. For the Hopfield model in a transverse field, we focused on the low memory loading regime and studied specific realizations of patterns. We first computed the static free energy per spin $f_s^{\mathrm{HM}}$ numerically and without recourse to self-averaging. We then computed the gaussian fluctuations to the ground-state energy, the inter-pattern energy gap in the condensed phase, and the excitation gap in the paramagnetic phase. 

In the path integrals of short-range interactive systems such as one-dimensional chains or two-dimensional lattices, using the multi-dimensional Hubbard-Stratonovich transformation $\exp(\frac{1}{2}\vec{\sigma}\cdot\textbf{J}\cdot\vec{\sigma})=\mathrm{const.}\int d\vec{m} \exp(-\frac{1}{2}\vec{m}\cdot\textbf{J}^{-1}\cdot\vec{m}-\vec{m}\cdot\vec{\sigma})$ to uncouple the interaction $\mathbf{J}$ between spins $\vec{\sigma}$ leads to single spin trace terms similar to Eq. (\ref{eq.sec_I.05}). However, the Hubbard-Stratonovich fields $\vec{m}$ in this case are generally not order parameters so Eq. (\ref{eq.sec_I.04}) and our method for expanding the trace perturbatively may not be applicable. For general spin traces like these, Lie-algebraic based methods for evaluating the time-ordered exponentials have been developed for lattice systems\cite{Galitski11,Ringel13}.

In this work, we considered the Hopfield model at low memory loading where it is possible to study specific realizations of disorder. However, for other types of disordered system, such as the Sherrington-Kirkpatrick model\cite{Thirumalai89} or the Hopfield model at high memory loading\cite{Nishimori96}, the methods used here for low memory loading may not be applicable. In order to apply the dynamical path integral approach we proposed here, it might then be necessary to first average over the disorder using the replica trick\cite{Thirumalai89}. Note, however, that once disorder-averaging has been performed, the finite-size corrections to the partition function and the energy gap being calculated will also be averaged quantities. Important features particular to specific realizations of disorder, such as quantum annealing bottlenecks within the spin glass phase\cite{Knysh16}, will also be lost. Nevertheless, such averaged quantities can also reveal much about the complexities of disordered models, and it would be interesting to extend our approach to the investigation of such systems in a future work. 

Lastly, we comment on the physical significance of the perturbative expansion of the trace $\mathcal{T}$. As mentioned in the Introduction, the formulation of $Z$ in the adiabatic representation by Knysh\cite{Knysh16} yielded a kinetic term in place of $\mathcal{T}$. The expansion presented in this paper can therefore be interpreted as the dominant terms of the kinetic energy. One may view our expansion as a somewhat tedious way to obtain what is only an approximation of the kinetic energy. However, one possible benefit of our approach is that the formulae [e.g. Eqs. (\ref{eq.sec_II.12})-(\ref{eq.sec_II.16})] can be reapplied on another suitable model simply by a change of Hamiltonian parameters. The calculations that ensue are quite undemanding from a computational point of view. The gaussian fluctuations can be calculated analytically; for the next order terms, the numerical evaluation of the double summations in Appendix \ref{app.C.derivation z412z312} is also very rapid. We think that this might be helpful when analyzing disordered models, where traditional methods such as exact diagonalization can be quite costly computationally speaking.

\begin{acknowledgements}
This work was partly supported by the Biomedical Research Council of A*STAR (Agency for Science, Technology and Research), Singapore. 
\end{acknowledgements}

\appendix

\section{Calculation of $|\sigma^{(r)}(t)\rangle$ and $\mathcal{T}^{(r)}$ using time-dependent perturbation theory}
\label{app.A.TDPT}

We derive $|\sigma^{(r)}(t)\rangle$ of Eq. (\ref{eq.sec_II.08}) and $\mathcal{T}^{(r)}$ of Eq. (\ref{eq.sec_II.10}) using time-dependent perturbation theory. For details on the latter, see \cite{Ballentine98}.

Let $\varepsilon_{\pm}=\pm \varepsilon$ and $|\pm\rangle$ denote, respectively, the eigenvalues and eigenvectors of $\mathcal{H}_s$. With $\{|+\rangle,|-\rangle\}$ as basis, expand 
\begin{equation}
|\sigma(t)\rangle
=
\sum_{n=\pm}
\sigma_n(t)e^{\varepsilon_n t}|n\rangle.
\label{eq.app_A.01}
\end{equation}
In Eq. (\ref{eq.app_A.01}), $e^{\varepsilon_n t}$ takes care of the time-dependence due to $\mathcal{H}_s$ while $\sigma_n(t)$ takes care of that due to $\lambda\mathcal{H}_d(t)$. The objective is to solve for $\sigma_n(t)$. Substituting Eq. (\ref{eq.app_A.01}) into Eq. (\ref{eq.sec_I.06}), we have
\begin{equation}
\frac{d\sigma_m(t)}{dt}
=
\lambda
\sum_{n=\pm}
\sigma_n(t)
e^{(\varepsilon_n -\varepsilon_m )t}
\langle m|\mathcal{H}_d(t)|n\rangle.
\label{eq.app_A.02}
\end{equation}
Expand $\sigma_n(t)$ in powers of $\lambda$
\begin{equation}
\sigma_n(t)
=
\sigma^{(0)}_n
+
\lambda
\sigma^{(1)}_n(t)
+
\lambda^2
\sigma^{(2)}_n(t)
+
\cdots,
\label{eq.app_A.03}
\end{equation}
where $\sigma^{(r)}_n(t)$ denotes the $r$th-order approximation of $\sigma_n(t)$. $\sigma^{(0)}_n$ is independent of time and determined by the initial condition $|\sigma(0)\rangle$. For $r\ge1$, $\sigma_n^{(r)}(0)=0$. With Eq. (\ref{eq.app_A.03}), Eq. (\ref{eq.sec_II.08}) becomes
\begin{equation}
|\sigma(t)\rangle
=
\left(
\begin{array}{c}
\sigma^{(0)}_+ e^{\varepsilon t}\\
\sigma^{(0)}_- e^{-\varepsilon t}\\
\end{array}
\right)
+
\lambda
\left(
\begin{array}{c}
\sigma^{(1)}_+(t) e^{\varepsilon t}\\
\sigma^{(1)}_-(t) e^{-\varepsilon t}\\
\end{array}
\right)
+
\lambda^2
\left(
\begin{array}{c}
\sigma^{(2)}_+(t) e^{\varepsilon t}\\
\sigma^{(2)}_-(t) e^{-\varepsilon t}\\
\end{array}
\right)
+
\cdots
.
\label{eq.app_A.04}
\end{equation}
Substituting Eq. (\ref{eq.app_A.03}) into Eq. (\ref{eq.app_A.02}) and collecting together the same powers of $\lambda$, one obtains the recursive relation
\begin{equation}
\frac{d}{dt}
\left(
\begin{array}{c}
\sigma_+^{(r+1)}(t) \\
\sigma_-^{(r+1)}(t) \\
\end{array}
\right)
=
2\beta J
\left(
\begin{array}{cc}
\alpha m_d(t) & \gamma m_d(t) e^{-2\varepsilon t} \\
\gamma  m_d(t) e^{2\varepsilon t}& -\alpha m_d(t)\\
\end{array}
\right)
\left(
\begin{array}{c}
\sigma_+^{(r)}(t) \\
\sigma_-^{(r)}(t) \\
\end{array}
\right).
\label{eq.app_A.05}
\end{equation}
Starting from the lowest-order coefficients $\sigma_n^{(0)}$, the $(r+1)$th-order coefficients are obtained recursively by integrating the $r$th-order ones. 

As an example, we calculate $|\sigma^{(1)}(t)\rangle$ and $\mathcal{T}^{(1)}$. Integrating Eq. (\ref{eq.app_A.05}), we have
\begin{equation}
\left(
\begin{array}{c}
\sigma_+^{(1)}(t) \\
\sigma_-^{(1)}(t) \\
\end{array}
\right)
=
2\beta J
\left(
\begin{array}{cc}
\alpha \int_0^t dt^{\prime} m_d(t^{\prime}) & \gamma \int_0^t dt^{\prime}m_d(t^{\prime}) e^{-2\varepsilon t^{\prime}}\\
\gamma \int_0^t dt^{\prime} m_d(t^{\prime}) e^{2\varepsilon t^{\prime}}& -\alpha \int_0^t dt^{\prime} m_d(t^{\prime})\\
\end{array}
\right)
\left(
\begin{array}{c}
\sigma_+^{(0)} \\
\sigma_-^{(0)} \\
\end{array}
\right)
.
\label{eq.app_A.06}
\end{equation}
To distinguish between the two basis kets $|\sigma(t)\rangle$, we denote $\sigma_n(t)=a_n(t)$ for $|\sigma(t)\rangle=|+1(t)\rangle$ and $\sigma_n(t)=b_n(t)$ for $|\sigma(t)\rangle=|-1(t)\rangle$. The summand of $\mathcal{T}^{(1)}$ with $\sigma=+1$ is
\begin{equation}
\langle +1(0) | +1^{(1)}(1)\rangle
=
\left(
\begin{array}{cc}
 a_+^{(0)}  &  a_-^{(0)} \\
\end{array}
\right)
\left(
\begin{array}{c}
a_+^{(1)}(1) e^{\varepsilon} \\
a_-^{(1)}(1) e^{-\varepsilon}\\
\end{array}
\right)
,
\label{eq.app_A.07}
\end{equation}
where $a_n^{(1)}(1)$ is given by Eq. (\ref{eq.app_A.06}). Summing with $\langle -1(0) | -1^{(1)}(1)\rangle$, and then using the orthonormal properties of the eigenvectors of $\sigma^z$, we obtain Eq. (\ref{eq.sec_II.13}). Higher-order terms are calculated similarly.

\section{Summary of static approximation results for the ferromagnetic model}
\label{app.B.summary static approx}

Making the static approximation $m(t)\approx m_s$, $\mathcal{T}\approx\mathcal{T}^{(0)}$ given by Eq. (\ref{eq.sec_II.12}). $Z$ becomes
\begin{equation} 
Z
\propto
\exp
\left(
-\beta N f_s
\right),
\label{eq.app_B.01}
\end{equation} 
where
\begin{equation} 
f_s=Jm_s^2 -\frac{1}{\beta} \ln 2 \cosh \varepsilon,
\label{eq.app_B.02}
\end{equation} 
is the free energy per spin. $m_s$ and $f_s$ are determined self-consistently using $\partial f_s/\partial m_s=0$. In the limit $\beta\rightarrow \infty$, we have 
\begin{equation}
m_s
=
\left\{
\begin{array}{ccc}
0 & \mathrm{for} & \Gamma \ge2J, \\
\pm\sqrt{1-\left(\frac{\Gamma}{2J}\right)^2} & \mathrm{for} & \Gamma<2J, \\
\end{array}
\right. 
\label{eq.app_B.03}
\end{equation}
and
\begin{equation}
N f_s
=
\left\{
\begin{array}{ccc}
-N \Gamma  & \mathrm{for} & \Gamma \ge2J, \\
-N \frac{(2J)^2+\Gamma^2}{4J}  & \mathrm{for} & \Gamma <2J. \\
\end{array}
\right.
\label{eq.app_B.04}
\end{equation}

\section{Derivation of $z_4^1, z_4^2, z_3^1$, and $z_3^2$ in Eq. (\ref{eq.sec_III.03})}
\label{app.C.derivation z412z312}

In deriving the $N^{-1}$ term of $E_0$, we note that two things help simplify the calculations. Firstly, when integrating over $V_4$ and $\frac{1}{2}(V_3)^2$ with the gaussian $e^{-\sum_n g_nc_nc_{-n}}$, `cross terms' such as $c_nc_{m}$ vanish. Secondly, we keep only those terms that do not vanish in the limit $\beta\rightarrow\infty$ upon inserting $Z$ into Eq. (\ref{eq.sec_I.02}). The results are
\begin{align}
z_4^1
&
=
(2\beta J)^2 8\varepsilon \gamma^4 \tanh\varepsilon
\left[
\sum_{n=1}^{\infty}
\frac{1}{g_n[n]}
\right]
\left[
\sum_{n=1}^{\infty}
\frac{1}{g_n[n]}
-
4(2\varepsilon)^2
\sum_{n=1}^{\infty}
\frac{1}{g_n[n]^2}
\right],
\label{eq.app_C.01}\\[15pt]
z_4^2
&
=
(2\beta J)^2 16\varepsilon\alpha^2\gamma^2\tanh\varepsilon
\left[
\left(
\sum_{n=1}^{\infty}
\frac{1}{g_n [n]}
\right)^2
+
(2\varepsilon)^2
\sum_{n=1}^{\infty}
\sum_{m=1}^{\infty}
\left(
\frac{1}{g_n [n]}
+
\frac{1}{g_m [m]}
\right)
\frac{1}{[n+m][n-m]}
\right.
\nonumber\\
&
+
\left.
(2\varepsilon)^2
\left(
2g
+
3(2\varepsilon)^2
\right)
\sum_{n=1}^{\infty}
\sum_{m=1}^{\infty}
\frac{1}{g_n g_m[n][m][n+m][n-m]}
\right],
\label{eq.app_C.02}\\[15pt]
z_3^1
&
=
\frac{(2\beta J)^4 4\alpha^2 \gamma^4 \tanh^2\varepsilon}{\beta J g_0}
\left[
\sum_{n=1}^{\infty}
\frac{1}{g_n[n]}
+
8\varepsilon^2
\sum_{n=1}^{\infty}
\frac{1}{g_n[n]^2}
\right]^2
,
\label{eq.app_C.03}\\[15pt]
z_3^2
&
=
(2\beta J)^3 8(2\varepsilon)^4\alpha^2\gamma^4\tanh^2\varepsilon
\sum_{n=1}^{\infty}\sum_{m=1}^{\infty}
\frac{1}{g_{n+m}g_ng_m [n+m]^2 }
\left[
\frac{1}{[n]}
+
\frac{1}{[m]}
+
\frac{(2\pi n)(2\pi m)+(2\varepsilon)^2}{[n][m]}
\right]^2,
\label{eq.app_C.04}
\end{align}
where $[n]=(2\pi n)^2+(2\varepsilon)^2$. The term $z_4^1$ stems from adding $-\frac{1}{2}(L_2)^2$ to the term involving $e^{\varepsilon}M_{-+-+}+e^{-\varepsilon}M_{+-+-}$ of $\mathcal{T}^{(4)}$, and $z_4^2$ stems from the remaining terms of $\mathcal{T}^{(4)}$. The term $z_3^2$ stems from $\frac{1}{2}(L_3)^2$, and $z_3^1$ stems from the remaining terms of $\frac{1}{2}(V_3)^2$. Along the way, the formula
\begin{equation}
\sum_{n=1}^{\infty}
\frac{1}{z_1 n^2 + z_2}
=
-\frac{1}{2z_2}
+
\frac{\pi}{2\sqrt{z_1z_2}}
\,
\mathrm{coth}
\left(
\pi\sqrt{\frac{z_2}{z_1}}
\right)
,
\label{eq.app_C.05}
\end{equation}
is used to evaluate some of the summations that appear and to check their powers of $\beta$. 

For $z_4^1$ and $z_3^1$, using partial fractions to simplify the summands and then using Eq. (\ref{eq.app_C.05}), one further obtains
\begin{align}
z_4^1
&
=
\frac{(2\beta J)^2(2\varepsilon)^2 \gamma^4}{g}
\left[
\frac{1}{\sqrt{(2\varepsilon)^2-g}}
+
\frac{1}{(2\varepsilon)^2-g}
\left(
\frac{g}{8\varepsilon}-2\varepsilon
\right)
\right],
\label{eq.app_C.06}\\[10pt]
z_3^1
&
=
\frac{m_s^2(2\beta J)^6\gamma^4}{\beta J g_0}
\left[
\frac{4}{g^2}
+
\frac{1}{(2\varepsilon)^2-g}
\left[
\frac{1}{(2\varepsilon)^2}
+
\frac{4}{g}
+
\left(\frac{4\varepsilon}{g}\right)^2
\right]
-
\frac{2}{g\sqrt{(2\varepsilon)^2-g}}
\left[
\frac{1}{\varepsilon}
+
\frac{8\varepsilon}{g}
\right]
\right]
,
\label{eq.app_C.07}
\end{align}
where terms of order $\beta^0$ and smaller have been dropped. In the ferromagnetic phase, they become
\begin{align}
z_4^1
=
&
\,
\beta
\left[
\frac{\Gamma^2}{2\sqrt{(2J)^2-\Gamma^2}}
-
\frac{\Gamma^2[(4J)^2-\Gamma^2]}{16J[(2J)^2-\Gamma^2]}
\right]
\,\,\,\,
\mathrm{for}
\,\,
\Gamma<2J
,
\label{eq.app_C.08}\\
z_3^1
=
&
\,
\beta 
\left[ 
J
+ 
\frac{J\Gamma^4 }{(2J)^2-\Gamma^2}\left[\frac{1}{16J^2} + \frac{1}{\Gamma^2} + \left(\frac{2J}{\Gamma^2}\right)^2 \right]
-
\frac{J\Gamma^2}{\sqrt{(2J)^2-\Gamma^2}}\left[\frac{1}{2J} + \frac{4J}{\Gamma^2}\right]  
\right]
\,\,\,\,
\mathrm{for}
\,\,
\Gamma<2J
.
\label{eq.app_C.09}
\end{align}
Lastly, we comment on the numerical calculation of the double summations appearing in Eqs. (\ref{eq.app_C.02}) and (\ref{eq.app_C.04}). A double sum $\Xi$ is computed for several large values of $\beta$ while keeping all the other parameters fixed. Fitting a straight line to $\ln\Xi=-s_1\ln\beta+s_2$, we determined $s_1$ and $s_2$. This gives $\Xi=e^{s_2}\cdot\beta^{-s_1}$, the asymptotic form of $\Xi$ as $\beta\rightarrow\infty$. The term $\beta^{-s_1}$ will ultimately be cancelled by other $\beta$, leaving $e^{s_2}$ as the contribution to $z_4^2$ or $z_3^2$.

\section{Calculation of $Z_{A_z}$ in the paramagnetic phase}
\label{app.D.Calculation of ZAz}

In the paramagnetic phase,  we have
\begin{equation}
\frac{\mathcal{T}_0 \mathcal{T}_{3z} + (N-1)\mathcal{T}_{1z} \mathcal{T}_{2z} }{(\mathcal{T}_0)^2}
=
\mathrm{sech}4\varepsilon
+
(g')^2
\sum_{n=-\infty}^{\infty}
c_n c_{-n}
\frac{(-1)^n }{[(\pi n)^2+(4\varepsilon)^2]^2}
+
O(N^{-1/2})
,
\label{eq.app_D.01}
\end{equation}
where we have dropped the cross terms. Performing the gaussian integrals, simplifying the summand of the series using partial fractions, and using
\begin{equation}
\sum_{n=1}^{\infty}
\frac{(-1)^n}{z_1 n^2+z_2}
=
-\frac{1}{2z_2}
+
\frac{\pi}{2\sqrt{z_1 z_2}}
\mathrm{cosech}
\left(
\pi\sqrt{\frac{z_2}{z_1}}
\right)
,
\label{eq.app_D.02}
\end{equation}
we obtain
\begin{equation}
Z_{A_z}
\stackrel{\mathrm{l.a.}}{\longrightarrow}
e^{-\beta N f'_s}
\,
\frac{\sinh4\varepsilon \tanh4\varepsilon}{\sinh^2 4\sqrt{\varepsilon^2-\frac{g'}{16}}}
\,
\frac{\varepsilon}{\sqrt{\varepsilon^2-\frac{g'}{16}}}
.
\label{eq.app_D.03}
\end{equation}
Eq. (\ref{eq.app_D.03}) is the same as Eq. (\ref{eq.sec_VC.03}) except for $\frac{\varepsilon}{\sqrt{\varepsilon^2-\frac{g'}{16}}}$ which is irrelevant in the limit $\beta\rightarrow\infty$.

\end{document}